\documentclass[12pt,preprint]{aastex62}
\shorttitle{Black hole spin evolution in collapsars}
\shortauthors{Janiuk et al.}

\begin{document}

\title{Accretion in a dynamical spacetime and
 spinning up of the black hole in the gamma ray burst central engine}

\author{Agnieszka Janiuk}
\affil{
Center for Theoretical Physics, Polish Academy of Sciences, 
Al. Lotnikow 32/46, 
02-668 Warsaw, Poland}
\email{agnes@cft.edu.pl}
\author{Petra Sukova}
\affil{Astronomical Institute, Czech Academy of Sciences, 
Bo\v{c}n\'i II 1401, 141 00 Prague, Czech Republic}
\author{Ishika Palit}
\affil{
Center for Theoretical Physics, Polish Academy of Sciences, 
Al. Lotnikow 32/46, 
02-668 Warsaw, Poland}

\begin{abstract}

We compute the evolution of a quasi-spherical, slowly rotating
accretion flow around a black hole, whose mass and spin
evolve adequately to the mass-energy transfer through the
horizon.
Our model is relevant for the central engine driving a long
gamma ray burst, that originates from the collapse of a massive star.
The computations of a GRB engine 
in a dynamically evolving spacetime metric are important specifically 
due to the transient nature of the event, in which a huge amount of mass is accreted 
and changes the fundamental black hole parameters, its mass and spin,
during the process. 
We discuss the results in the context of angular momentum magnitude of the 
collapsing star.
We also study the possible formation and evolution of shocks in the envelope, 
which may temporarily affect accretion.
Our results are important for the limitations on the mass and spin 
range of black holes detected independently by
electromagnetic observations of GRBs and gravitational waves.
We speculate on the possible constraints for the final masses and spins of 
these astrophysical black holes. It is shown that the most massive BHs were rahter not formed in a powerful GRB explosion
 if the cores of their progenitors were only weakly rotating.

\end{abstract}
\keywords{accretion; black hole physics; gamma ray bursts:long; hydrodynamics}

\section{Introduction}

The collapsar model was originally proposed to describe the long-duration gamma-ray bursts
\citep{woosley1993, paczynski98}. 
In this scenario, the total energetics of explosion is consistent with
the amount of total binding energy of the
 progenitor star, while
the duration of the event, the time variability of the prompt phase, 
details of the afterglow emission, observations of host galaxies, and statistical
studies in connection with supernovae, may give further insight to 
the physics of collapsing massive stars \citep{podsiadlowski04, crowther2007}.

In general, all GRBs are presumably powered by accretion of a rotationally supported torus onto a newly born black hole. The hole itself should also fastly 
rotate, in order to efficiently transport the power to the remote jets. 
Both Blandford-Znajek process and neutrino anihillation may act as the source of this power, and in these two mechanisms their efficiency is similarly scaled with the black hole spin \citep{McKinneyGammie2004, liu2015, janiuk2017}. The process is 
further mitigated by the magnetic fields. 
For long GRBs, the rotation of the progenitor star is a key property in order to support accretion over relatively long activity periods, and also to sustain the rotation of the black hole \citep{janiukproga2008, moderski, bkomis08}.

In 2015, LIGO first discovered gravitational waves from a binary black hole merger, marking the beginning of a new era in astrophysics \citep{abbott2016}. This revolutionary discovery challenged our understanding of compact objects, including the formation of very heavy black holes from the direct collapse of massive stars \citep{spera2015}. 
The process of collapse inevitably involves feeding the new black hole with mass and angular momentum. Because rotation of the hole is essential 
for the occurrence of GRB, the open question remains about how fast the 
collapsed black hole of a given final mass may rotate (see, e.g. \citet{lopez2010, batta2016}). Or, in other words, 
whether the black holes as heavy as those found to date by LIGO (GW150914, GW151226, LVT151012) can also posses a large spin, and hence they could have been
in the past the 
progenitors of extremely energetic or ultra-long GRB explosions.
That could in principle be possible if the black hole had been spun up 
by its companion in a close binary system \citep{bkomis10, bbh2013}.
The black hole spin constraints are however not directly available from the 
gravitational waveform analysis, which gives only the mass-weighted projection 
of their values, i.e., so called 'effective spin' of the binary black hole merger \citep{cutler1994}. 

Our study of the star collapse accounts for the simultaneous 
numerical constraint for the black hole
mass and spin. It may help obtain a more consistent picture of the 
 stellar mass black hole formation. This is particularly important
 in the view of currently available 
 and future multi-messenger observations, and the ongoing debate on the relevance
 of the electromagnetic observations for the measurement of the intrinsic black hole parameters
 \citep{pankov17}. 
In this work, we perform self-consistently a numerical simulation
of the collapse process and black hole growth, considering the full General Relativistic framework.
We account for the change of black hole mass and spin during the accretion 
of the mass-energy through the black hole horizon, and we evolve the spacetime 
as a sequence of Kerr solutions.
The stellar structure is described using
a slowly rotating, quasi-spherical flow, with the relativistic solution for the Bondi-Michel radial 
dependence of density, and specific angular momentum concentrated at the equator (\citet{janiukproga2008}, see also e.g. \citet{mach2018}).
Our simulations utilize the changing Kerr metric coefficients and black hole growth, which we have implemented within the HARM code (High Accuracy Relativistic Magnetohydrodynamics, \citet{gammie2003harm}).

Our study is aimed to probe the influence of the amount of angular momentum in the quasi-spherical flow during the star's collapse on the final black hole properties. We follow the assumption, that only the fast spinning black holes, accompanied by the
action of a rotationally supperted 'mini-disk' embedded in the collapsar envelope, may drive the long
GRB emission \citep{moderski}.
In addition, a possible existence and evolution of shock fronts is studied, which is a common feature in the low-angular momentum black hole accretion flows \citep{chakrabarti2001, das2002, sukova2015, sukova2017}. The occurence of
these shocks may affect accretion, lead to a transient variability, and other potentially observable effects
in long GRBs.
Finally, we speculate that our computations may help solve the puzzle of formation of the
most massive stellar black holes as observed by LIGO, and maybe put an independent constraint on their
possible maximum spin values.

\section{Evolution of the black hole mass and spin}
\label{sec:bhevol}
	
The model computations are based on the axisymmetric, general relativistic
MHD code HARM, described by \citet{gammie2003harm} and \citet{noble2006}.  
We investigate the evolution 
of an axisymmetric, slowly rotating, non-magnetized,
plasma accreting onto a black hole. Its key parameters (mass and spin) evolve with time
as for a gamma ray burst central engine, 
where the transient accretion episode results in a fastly changing physical 
conditions leading to the black hole growth.
We describe the collapsar evolution taking into account the changing Kerr metric due to the evolving mass and spin of BH
(see also \citet{karas} who studied the runaway instability of relativistic tori around accreting black holes).
In this sense, we significantly expand our previous calculations of the GRB central 
engine, which were conducted in a fixed background metric as suited for short GRBs \citep{janiuketal2013, janiuk2017}.

In the recent simulations of binary neutron star mergers, the evolution of 
spacetime is fully coupled with the evolution of matter until the stars merge and a compact remnant is formed. The massive compact remnant is then
dominating the spacetime evolution in the system, while the contribution of the
surrounding disk is not very large due to its small mass (see for instance
\citet{kastaun}).
Recent collapsar simulations, on the other hand, involve the conservation equation for the stress-energy tensor, including the fluid and radiation field,
and the metric evolution is followed through the
standard BSSN method. The black hole growth is however
not followed further after the
core-collapse \citep{ott2018}
or the black hole is found and diagnosed by means of the baryon mass enclosed inside a certain (i.e. Schwarzschild) radius. Its mass is then fixed and the hole is not rotating, so the metric is frozen
\citep{kuroda2018}.

Solving the full set of Einstein equations that describe the collapsar evolution in a dynamical spacetime for the whole duration of the event is a very complex task.
In our paper, we use a simple physical picture of the collapsing
massive star, and we start our simulation when the black hole has already formed.
We assume that its gravitational field determines the
subsequent spacetime evolution.
However, instead of using a stationary metric with a constant mass and spin of the black hole, and simply allowing the matter to accrete onto it,
we follow the sequence of stationary solutions with black hole mass and spin parameters updated by a very small value in each time step.
This approach assumes that the effect of the extended matter is still not
as dynamically important, as the effect of the black hole on the spacetime evolution (see e.g. \citet{semerak} and Fig. 1, right panel, in their paper, for the gravitational effect of a massive thin disc lying in the equatorial plane. Even in the case of a disc five times more massive than the black hole, the potential is only modestly altered. Moreover, this type of source, i.e. infinitesimally thin disc, is supposed to be a stronger source of gravitational field than a gaseous spherical envelope of the same mass). 
Therefore, we propose here a method which however is not meant by solving the field equations to describe the collapse including the envelope matter self-gravity, but still it gives a good approximation to this problem and accounts for changing Kerr metric.

In our present simulations, the evolution of the black hole mass $M$ and spin $J$ is computed according to 
the following equations \citep{gammie2004}:
\begin{equation}
\dot{J} \equiv \int d\theta d\phi\, \sqrt{-g}\, {T^{r}}_\phi, \nonumber \\
\label{eq:spin}
\end{equation}
\begin{equation}
\dot M = \dot{E} \equiv \int d\theta d\phi\, \sqrt{-g}\, {T^{r}}_t \nonumber \\
\label{eq:power}
\end{equation}
where the components of the stress energy tensor of the accreting matter, 
$T^{\mu}_{\nu}$, are integrated over the black hole horizon. 

The changing black hole spin and mass are subsequently 
affecting the spacetime metric. 
The mass growth is discretized and 
updated in every time step according to \citet{moderski}:
\begin{equation}
\Delta M  = {M_{\rm BH}^{\rm curr} \over M_{\rm BH}^{0}}-1,
\label{eq:dmc}
\end{equation}
where $M_{BH}^{0}$ is the initial mass of the black hole, and the current mass is given by
integration of the rest-mass flux over the horizon at every time-step:
\begin{equation}
  M^{\rm curr}_{\rm BH} = M_{\rm BH}^{i} = M_{\rm BH}^{i-1} + \int_{r=r_{\rm in}} dM_{\rm in} 2\pi d\theta \sqrt{-g} \Delta t,
\end{equation}
where
\begin{equation}
dM_{\rm in} = - \rho \frac{u^r}{u^t}.
\end{equation}

We subsequently update the six relevant coefficients of the $g_{\mu\nu}$ metric
in the Kerr-Schild form, 
which are dependent on the central mass, and are also sensitive to the 
spin change, namely:
\begin{eqnarray}
g_{tt}  =  -1 + 2(1+\Delta M) {r \over {r^{2}+a^{2} \cos^{2}\theta}}, \nonumber \\
g_{tr} =  2 (1+\Delta M) {r \over {r^{2}+a^{2} \cos^{2}\theta}}, \nonumber \\
g_{t \phi} = -2 (1+\Delta M) a r {\sin^{2}\theta \over {r^{2}+a^{2} \cos^{2}\theta}}, \nonumber \\
g_{r r} = 1 + 2 (1+\Delta M){r \over {r^{2}+a^{2} \cos^{2}\theta}}, \nonumber \\
g_{r \phi}  = -a \sin^{2}\theta (1+ 2 (1+\Delta M){r \over {r^{2}+a^{2} \cos^{2}\theta}},  \nonumber \\
g_{\phi \phi} = \sin^{2}\theta (r^{2}+a^{2}\cos^{2}\theta + \nonumber \\ 
+  a^{2}\sin^{2}\theta (1 + 2 (1+\Delta M) {r \over {r^{2}+a^{2} \cos^{2}\theta}} )).
\end{eqnarray}

The change of the spin parameter of the black hole is computed as:
\begin{equation}
a^{i} = a^{i-1} + ({\dot J \over M^{\rm curr}_{\rm BH}} -  {a^{i-1} \over M^{\rm curr}_{\rm BH} } {\dot E }) \Delta t.
\label{eq:spinevol}
\end{equation}
Note that the original HARM code works in dimensionless units, $G=c=M_{\rm BH}=1$, so they do not
appear in the metric coefficients explicitly.
The current mass of the black hole is given by 
${M_{BH}^{curr} \over M_{BH}^{0}}=1+\Delta M$.

The spin parameter $a$ here is not dimensionless, but it has the unit of $M_{\rm BH}$.
Also, the spatial and time coordinates are measured in the units of the black hole mass $M_{\rm BH}$. The dimensionless black hole spin, defined as $s=a/M=J/M^{2}$, is the one for which the Equation (\ref{eq:spin}) holds. Hence, by determining the derivative
of $a$, $da/dt = (ds/dt) M + s (dM/dt)$, we get the relation (\ref{eq:spinevol}) for the black hole spin evolution.

In the following sections, we consider a specific astrophysical scenario 
for the collapsing star and GRB progenitor, so that the physical units will be chosen.
We scale our calculations in a such way that the mass of the star is in the range of a
few tens of Solar mass. In order to compute the mass increase properly, we 
 evolve the ratio $\Delta M$. We consider the 
initial black hole mass of a certain value,  $M_{BH}^{0} = 3 M_{\odot}$, which is on
the order of an initial Iron core mass in the collapsing star.

\section{Initial conditions and dynamical model}

Our initial conditions are considering the case relevant to the 
long gamma ray bursts.
The initial condition is prescribed as a slowly rotating 
spherical cloud of matter, given by the Bondi solution in the Kerr metric,
supplied with a small angular momentum.
The Bondi flow extends from the outer boundary of the grid, 
typically located at a 
thousand gravitational radii, to the black hole horizon, 
and 
is parameterized by the location of the sonic radius.
We integrate the initial density and radial velocity profiles upwards and downwards from
the sonic radius.

The angular momentum is prescribed as a fraction of the {\it critical} angular momentum, which is the value at the circularisation radius for a test particle orbiting a black hole.
Here, the energy and angular momentum are given by:
\begin{equation}
\varepsilon = {{1 - 2/r +a/r^{3/2}} \over {\sqrt{1-3/r + 2 a /r^{3/2} }} }
\end{equation}
\begin{equation}
l = {{r^{1/2} - 2 a/r + a^{2}/r^{3/2}} \over {\sqrt{1-3/r + 2 a /r^{3/2} }} }
\end{equation}

We start all simulations in this work with non-spinning black holes, and in most cases we 
take the value of the angular momentum at the last stable orbit for a non spinning black hole (i.e., at $r=6 r_{g}$)which is then mapped onto the whole radial grid.
The angular velocity in the Boyer-Lindquist coordinates is then given by:
\begin{equation}
u^{\phi} = g^{t\phi}(-\varepsilon) + g^{\phi\phi} l
\label{eq:rotation}
\end{equation}
where $g^{t\phi}=-2ar/(\Sigma\Delta)$ and $g^{\phi\phi}=(\Delta-a^{2}\sin^{2}\theta)/(\Sigma\Delta\sin^{2}\theta)$, with $\Sigma = r^{2}+a^{2}\cos^{2}\theta$ and
$\Delta = r^{2}-2r+a^{2}$.
In addition, we want to scale rotation of the cloud (collapsing envelope) to have a maximum at the
equatorial plane, and tend to zero at the poles, so that we introduce an additional factor of $\sin^{2}\theta$. Finally, we scale our models with a
parameter defining a ratio between our collapsar's angular momentum and the above (critical)
angular momentum value at the circular orbit. Therefore we use:
\begin{equation}
l_{\rm spec} = S |{u_{\phi}\over u_{t}}| \sin^{2}\theta
\end{equation}
with $u^{\phi}$ defined in Eq. \ref{eq:rotation} and $S \equiv l/l_{\rm crit}$ being a model parameter, 
in principle larger, or smaller than unity.

The initial black hole mass is assumed fixed, 
and equal to $M_{\rm BH}=3 M_{\odot}$. 
The total
mass of the surrounding gas is computed taking into account the 
density scaling unit, which
gives the cloud of $M_{\rm cloud} \approx 25 ~M_{\odot}$ contained within 
the Bondi sphere of the size $R_{\rm out}=1000 ~r_{\rm g}$.
This reflects the core collapse scenario for the central part of a massive evolved star, which is 
contained within some $5 \times 10^{8}$ cm, while the outer 
layers that contain mostly the light Hydrogen envelope are neglected in our simulations,
for practical reasons (cf. \citet{janiukproga2008}).
The sonic point is chosen to be a parameter of our initial configuration, and determines the density and velocity 
profile according to the transonic Bondi solution. In our models, the sonic radius is
initially located at $r_{s}=80 r_{g}$, well inside the computational domain.
The models assume weak rotation profile as described above, but 
neglect the magnetic fields. 
The equation of state adopted in our simulations is given by a 
$P=(\gamma-1)u$, where $P$ is the pressure and $u$ internal energy of the gas,
and we use the adiabatic index of $\gamma=4/3$. 

To follow the
evolution of the gas dynamics near a black hole we use a numerical MHD
code {\it HARM-2D} \cite{gammie2003harm, noble2006}. 
The original code was designed to solve
magnetohydrodynamic equations in the stationary metric around a black
hole. The code is written in a conservative, shock-capturing
scheme, and has a low numerical viscosity.

In this work, we modify the original
code to account for 
the change of the  background metric, according to the amount of mass
sinking under the horizon, and supplementing the black hole with mass
and energy-momentum.
The metric is updated in every dynamical time-step during the simulation. 
To speed up the computations and obtain better efficiency, we use our own parallelization scheme, based on 
Open-MP and optimized through the distribution of the processes among the physical system directions\footnote{Provided by the function \textit{MPI\_Dims\_create(nprocs, 2, dims)} using an appropriate divisibility algorithm}.

All models presented here have numerical resolution of the grid 256x256 points
in $r$ and $\theta$ directions.
The grid is logarithmic in radius and
condensed in polar direction towards the equatorial plane, 
as in \citet{gammie2003harm}.

\section{Results}\label{sec:results}

The full list of models and their parameters is given in the Table \ref{table:models}.
Note that the CL and SL, as well CT and ST runs, have the same values of
physical parameters, respectively. They differ between each other with respect to the final time
of the evolution. Still, information about the collapsar properties at shorter end time of the simulation
was important, because for some particular values of parameters the longer simulation runs were
leading to almost complete evacuation of the accreting cloud (at the level of numerical density 'floor').

\begin{table*}
\begin{center}
\caption{Summary of the models. Mass is given in the units of
  $M_{\odot}$, radius is in the units of $GM/c^{2}$. End time of the simulations is given in seconds, for a black hole mass $3 M_{\odot}$. Mass lost through the inner boundary is integrated over the time of the simulation, and given in the units of Solar mass. $M^{0}_{\rm BH}$ is equal to 3 $M_{\odot}$ in all runs, and $M^{0}_{\rm cloud}=25 M_{\odot}$, and $a^{0}=0$ initially. 
\label{table:models}
}
\begin{tabular}{ccccccccccc}
\hline
Model & $r_{c}$ & $S$ & $M^{end}_{\rm BH}$ & $t^{end}$ & $a^{\rm end}$ & $s^{\rm end}$ & Metric change &  $M_{\rm cloud}^{end}$ & $M_{in}$ & $M_{out}$ \\
\hline   
CL-04 & 6 & 0.4 &  13.53 & 0.8 & 0.89 & 0.19 & yes & 2.69  & 10.53 & $10^{-19}$ \\
CL-10 & 6 & 1.0 &  9.49  & 0.8  & 2.22& 0.70 & yes & 7.39 & 6.5 & $6\times10^{-4}$ \\
CL-14 & 6 & 1.4 &  4.52  & 0.8 & 1.21 & 0.80 & yes & 20.51 & 1.52 & 1.11 \\
\hline
SL-04 & 6 & 0.4 & 14.51 & 3.0 & 0.98 & 0.20 & yes & 0.0007 & 11.51 & 0.0004  \\ 
SL-10 & 6 & 1.0 & 11.88 & 3.0 & 2.53 & 0.64 & yes & 0.014  & 8.88 & 0.002  \\ 
SL-14 & 6 & 1.4 & 6.27 & 3.0 & 2.09  & 0.99 & yes & 9.98 & 3.27 & 1.82  \\ 
\hline
CT-04 & 6 & 0.4 & 3.0 & 0.8 & 0 & 0 & no & 16.61 & 5.09 & $10^{-16}$  \\ 
CT-10 & 6 & 1.0 & 3.0 & 0.8 & 0 & 0 & no & 19.77 & 2.26 & 0.52  \\ 
CT-14 & 6 & 1.4 & 3.0 & 0.8 & 0 & 0 & no & 20.97 & 1.31 & 1.54  \\ 
\hline
ST-04 & 6 & 0.4 & 3.0 & 3.0 & 0 & 0 & no & 2.34 & 13.72 & $10^{-16}$ \\
ST-10 & 6 & 1.0 & 3.0 & 3.0 & 0 & 0 & no & 8.05 & 7.73  & 0.51 \\
ST-14 & 6 & 1.4 & 3.0 & 3.0 & 0 & 0 & no & 13.82 & 4.50 & 2.36 \\
\hline
RL-04 & 10 & 0.4 & 14.52 & 3.0 & 1.07 & 0.22 & yes & 0.0007 & 11.52 & 0.005 \\ 
RL-10 & 10 & 1.0 & 6.38 & 3.0  & 2.12 & 0.99 & yes & 0.49  & 3.38  & 1.57  \\ 
RL-14 & 10 & 1.4 & 6.82 & 3.0 & 2.00 & 0.88 & yes & 7.81 & 3.82 & 4.27  \\ 
\hline
\end{tabular}
\end{center} 
\end{table*}

\subsection{Evolution of the flow for different rotation magnitudes}

\begin{figure}
\includegraphics[width=7cm]{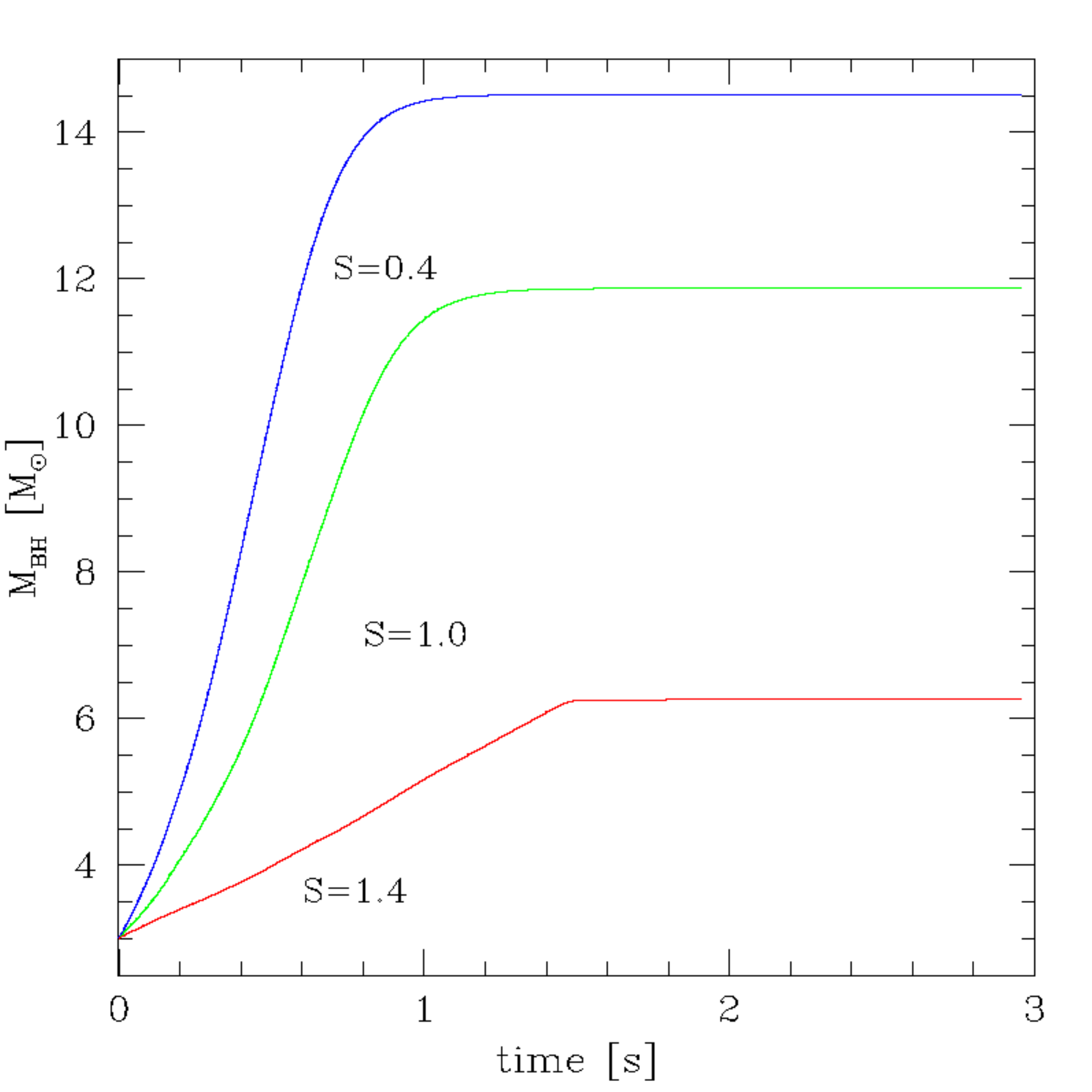}
\includegraphics[width=7cm]{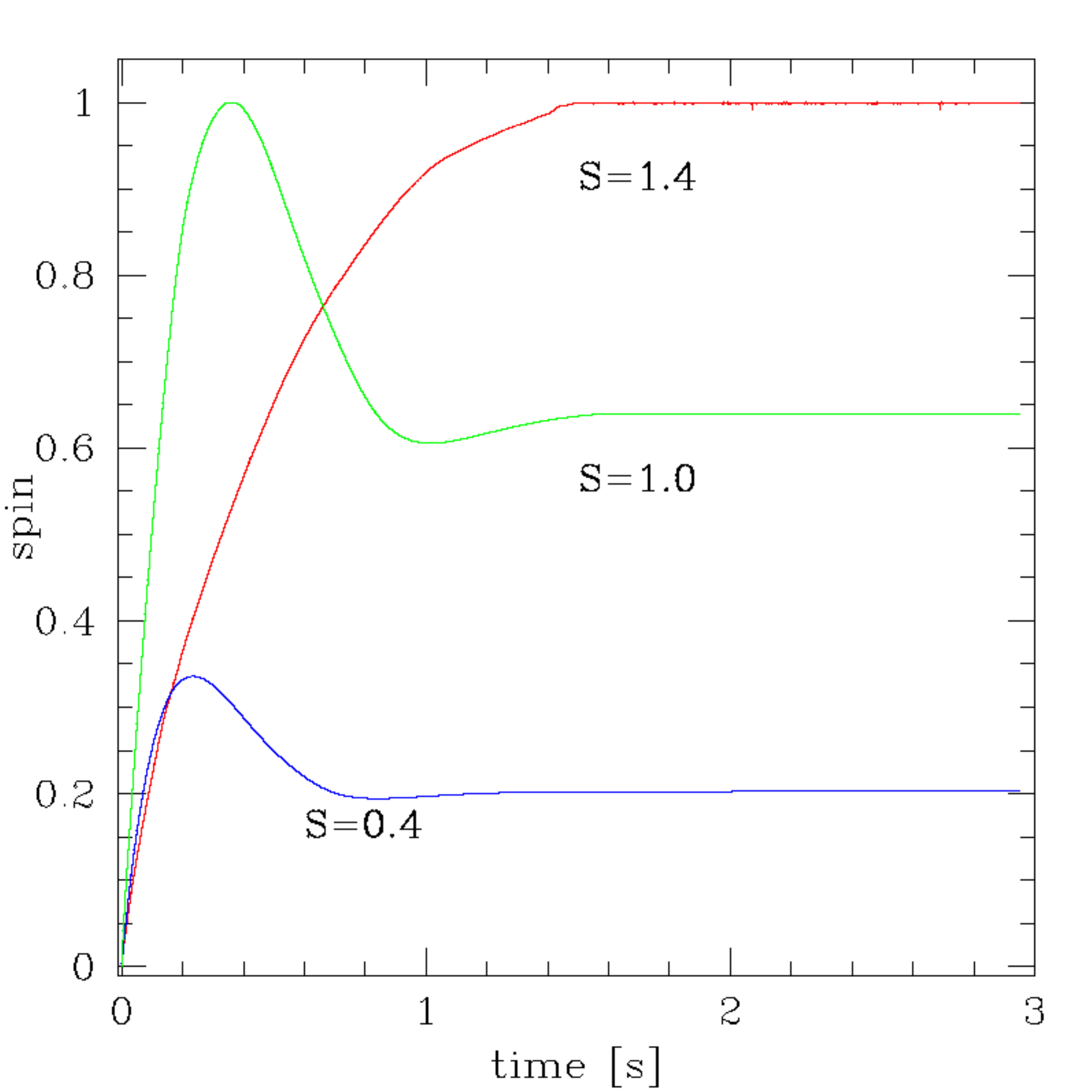}
\caption{Mass of the black hole as a function of time (top), and the dimensionless black hole spin evolution (bottom). The three lines show the model 
of a transonic 
cloud accreting onto black hole, with different values of the specific 
angular momentum. 
The time is given in seconds, and the final time of the evolution
 refers to $2\times10^{5}$ geometrical units, $GM/c^{3}$, where $M=M_{BH}^{0}=3 ~M_{\odot}$ is the initial black hole mass.
}
\label{fig:mass_spin_evol}
\end{figure}

The mass and spin of the black hole grow in time during the dynamical 
simulations, according to equations (\ref{eq:spin}) and (\ref{eq:power}). As expected, the fastest growth of the black hole mass occurs in case of the 
almost non-rotating flow, when the matter inside the cloud flows in 
supersonically through the horizon from all directions. 
The rotation halts the gas particles which are closer to the equatorial plane, and form a kind of 'mini-disk' there. In  models with higher $S$, the black hole mass grows only due to accretion of gas from the polar direction.

Figure \ref{fig:mass_spin_evol} presents the results of both mass and dimensionless spin evolution for the three models, with sub-critical, $S=0.4$, 
critical, $S=1.0$, and super-critical rotation, $S=1.4$. The critical value of $S$ means that the condition for the formation of a long-living mini-disk is satisfied.
Still, only in the model with $S=1.4$, the black hole spin grows, up to the value of $s=a/M \approx 0.99$ and remains high until the end of our simulation run.
In the sub-critical rotation case, the larger is the rotation parameter, the larger BH spin value is obtained at its maximum. For $S=1.0$ the temporarily achieved spin is as high as 0.99.
However, when the initial fast rotating bubble is accreted.
This bubble is accreted within  1 second. Then, the spin of the black hole decreases again and saturates at
$s=a/M \approx 0.65$.
For the slowest rotation model, $S=0.4$, the black hole spin at peak reaches value of 0.35, and then decreases to about 0.2.

We also checked if the maximum value of the spin is sensitive to the chosen circularisation radius (cf. runs RL in Table \ref{table:models}).
In most of our models, $r_{\rm c} = 6$ was chosen. 
We checked that when the 
value of $r_{\rm c}$ is larger, then 
the mini-disk size is larger, if the rotation is equal or larger than
critical.

The spin-up of the black hole occurs more effectively only
in the case only of critical rotation, and the final spin at the end of the run RL-10 is at maximum Kerr limit.
For super-critical rotation, the mass lost through the outer boundary due to the centrifugal force was in fact larger, than the mass accreted onto black hole. And since the rotationally supported mini-disk kept the material on the orbit and the increase in the BH spin per unit mass was lower, the final value of the dimensionless spin was in this run 
(RL-14) even smaller than for the generic model (SL-14).

Figure \ref{fig:mcloud} shows the accompanying change of the accreting cloud 
total mass, as integrated over the simulation volume.
The parameters of our model were chosen such that the initial mass was always 
equal to 25 $M_{\odot}$. The actual mass of the cloud decreases then due to 
accretion (inflow through the horizon), but also due to the outflow through the outer boundary. 
Final mass depends on how much matter is 
still halted in the mini-disk, or has not been accreted until the end of the 
simulation.
For sub-critical rotation, the cloud mass will asymptotically drop to zero, 
however the long simulation runs show, that a small blob remains even at 
late times in the equatorial plane, when the sub-sonic material comes to the 
innermost regions, and is decelerated by rotation.

\subsection{Comparison with the static metric models}

The mass and spin of the black hole, by definition, do not change
in the static metric models.
This setup, as is used for comparison and code testing, assumes that the mass is lost from the simulation volume through the inner (and also outer) boundary, however it does not contribute further to the black hole growth
nor metric change.

\begin{figure}
  \includegraphics[width=7cm]{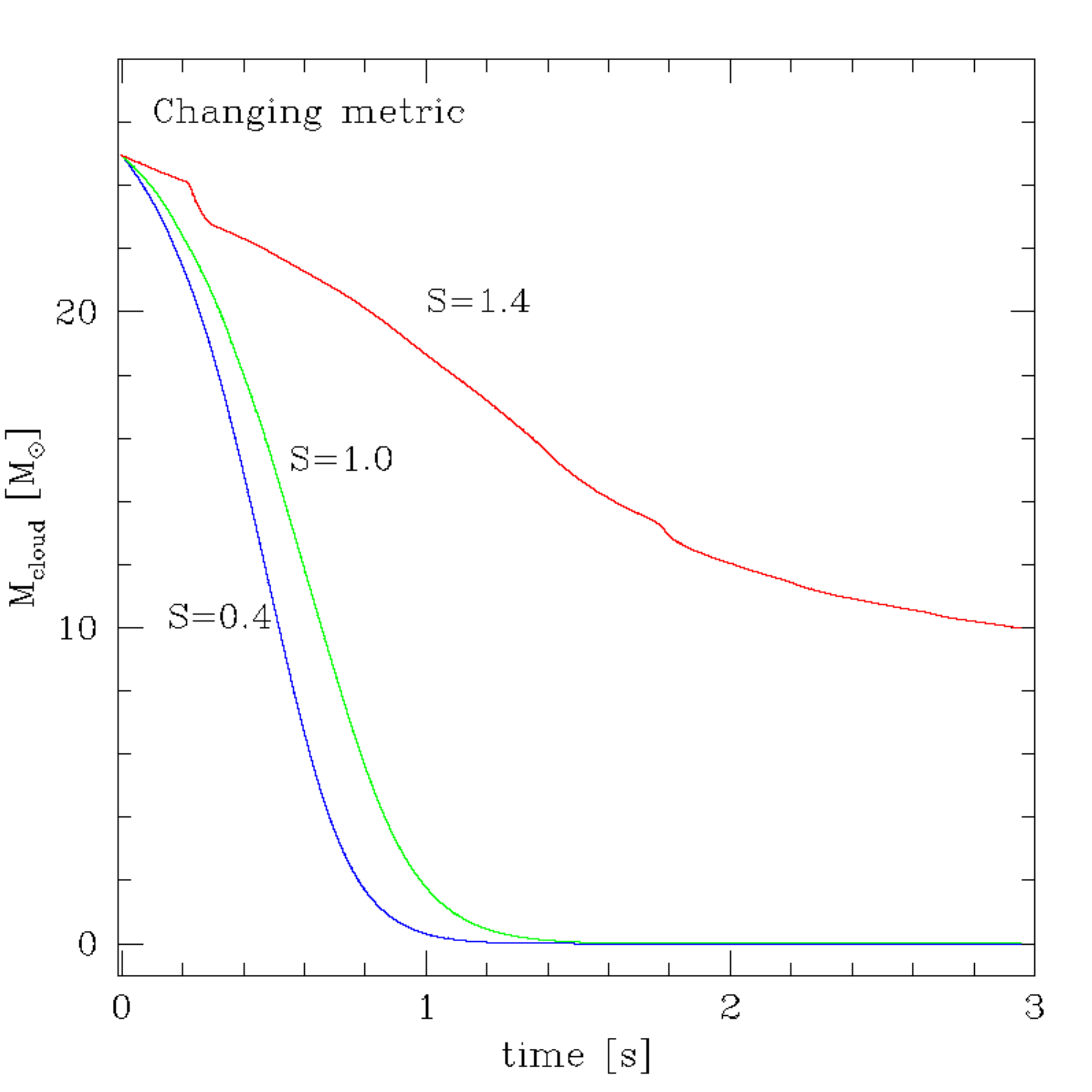}
  \includegraphics[width=7cm]{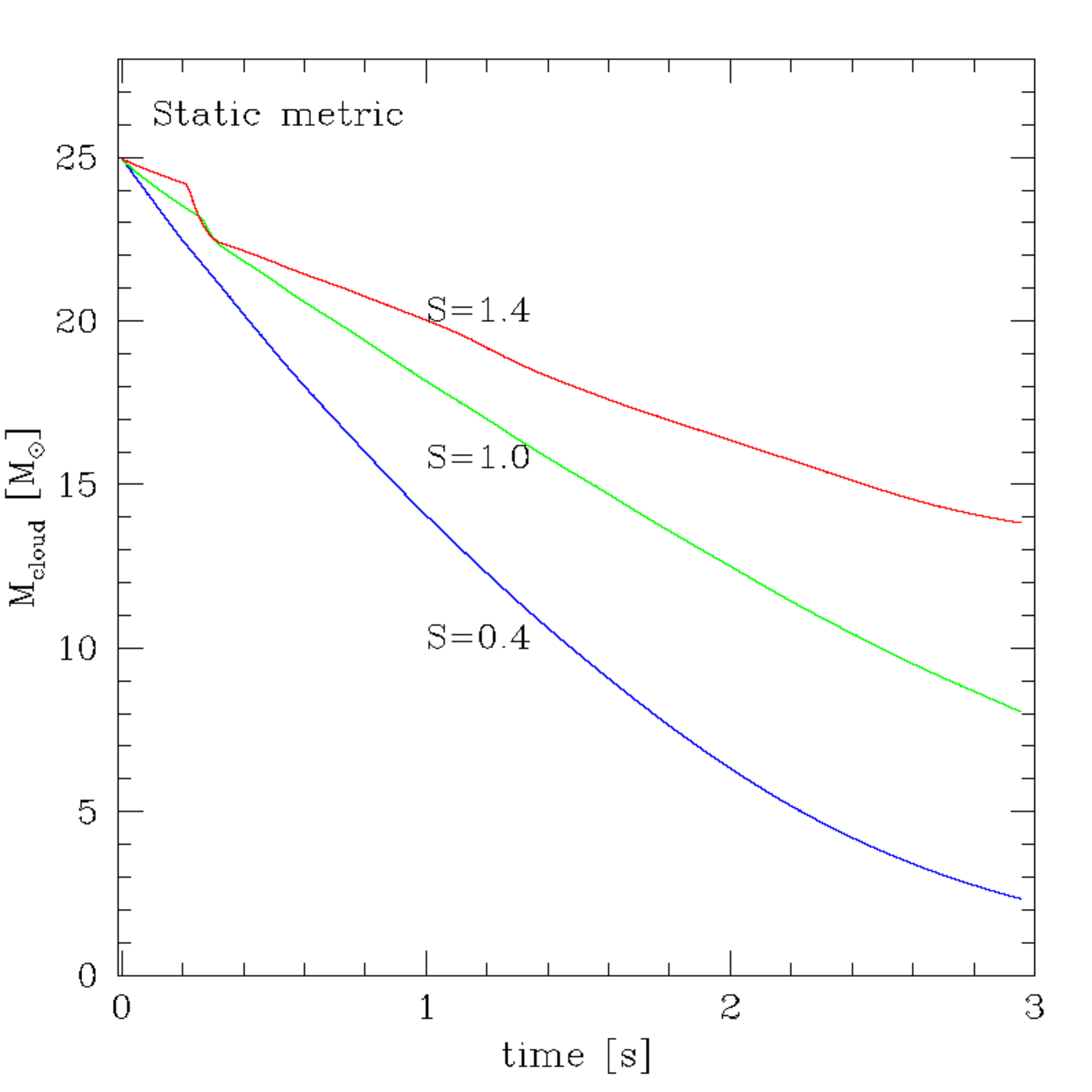}
\caption{Mass of the accreting cloud, contained within 1000 gravitational radii, as a function of time.
Three solid lines represent the rotation with critical angular momentum at $6 r_{g}$ (green), and for $0.4 ~l_{\rm crit}$ (blue), and for $1.4 ~l_{\rm crit}$ (red).
The bottom panel presents the results for a test run, where the
black hole mass and spin was not evolved ($M_{\rm BH}=3 ~M_{\odot}$ and $s=a=0$).
}
\label{fig:mcloud}
\end{figure}

The bottom panel of the Figure \ref{fig:mcloud} shows for comparison the evolution of the cloud mass in the stationary metric, when the black hole mass 
and spin are constant (cf. runs ST in Table \ref{table:models}).
As shown in the Figure \ref{fig:mcloud}, the static metric models result in much shallower decrease of the accreting cloud mass. The mass lost through $r_{in}$ was of moderate amount. The mass of the cloud was also lost through the outer boundary, due to the free outflow boundary condition.

The simulations with metric evolution in general show that much more mass is accreted onto black hole through the inner boundary, while the mass outflow
at the outer boundary is rather negligible (see Tab. \ref{table:models}).

We also stress, that the quantity which is conserved in the code, is the
comoving density, $D=\rho u^{t}$.
This is why the numbers in columns 4, 9, 10 and 11 do not add up to
the value of initial cloud mass ($25 M_{\odot}$)
We verified that the conservation scheme works within a good accuracy
in both setups. The volume integrated $D$ is conserved, taking
into account its outflow through the outer and inner boundaries, over the simulation time, with accuracy of $0.2\%$ and $0.3\%$ for the changing and
static metric runs, respectively.

In general, for static metric runs, one can notice a smoother decrease of mass, and slower accretion in case of sub-critically rotating flows. For the super-critical rotation, the mass decrease is similar in both cases, i.e. does not drop to zero neither in changing, nor in static metric simulations. The final mass of the cloud at time $t=3$ s is $M_{\rm c} \approx 10 M_{\odot}$, and  $M_{\rm c} \approx 14 M_{\odot}$, see runs SL-14 and ST-14, respectively.
We also note that a small jump occurs in these runs at $t=0.25$ s, see Fig. \ref{fig:mcloud}.
This is attributed to the shock accretion and formation of the mini-disk (see next Section), but the magnitude of this jump is not affected by the metric evolution.

\subsection{Mass accretion rate and 'mini-disk' formation}

\begin{figure}
 \includegraphics[width=7cm]{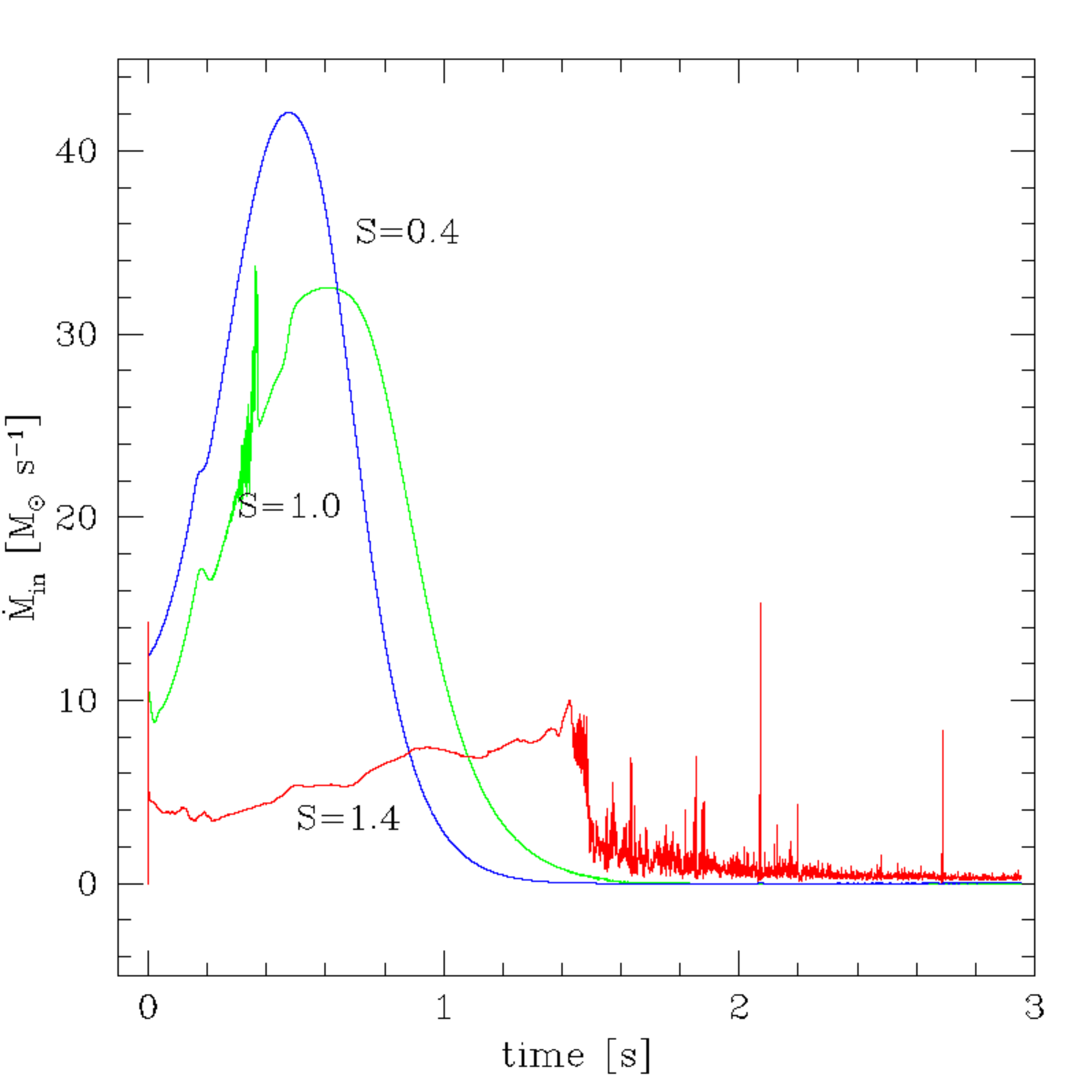}
\caption{Mass accretion rate as a function of time.
Three lines represent the rotation with critical angular momentum at 
$6 ~r_{g}$ (green), for $0.4 ~l_{\rm crit}$ (blue), and for $1.4 ~l_{\rm crit}$ (red).
Initial mass of the cloud was $25 ~M_{\odot}$ and the sonic point was located at $80 ~r_{\rm g}$.
}
\label{fig:accrate}
\end{figure}

In Figure~\ref{fig:accrate}, we show the time evolution of the mass accretion
rate onto black hole.
The temporary accretion rate is large, due to the fast supersonic
inflow, and for our parameters and physical units normalization
it is on the order of a few tens of $M_{\odot}$ s$^{-1}$.
 The accretion rate
 for the sub-critical rotation model is highest at peak, and steeply grows with time at the beginning of the simulation. The subsequent decrease of the accretion rate occurs when there is almost no matter left and the cloud empties.
 The flow here is highly supersonic and this is why the velocity of the inflow
 results in the rapid evacuation of the cloud material, and growth of the black hole.
In case of critical rotation, an additional effect is temporary flickering of the accretion rate around  $t\sim$ 0.37s ($25,000$ M \footnote{Due to the choice $M^0_{\rm BH} = 3M_\odot$ the conversion from geometrical units to physical units is such that $10, 000$ M corresponds to ca. 0.148 s.}), which is connected with the shock bubble vertical oscillations.
For supercritical rotation, the accretion rate is quite stable and constant with time during the shorter simulation, and is at the level of 5
$ M_{\odot}$ s$^{-1}$. 
For longer time, the accretion rate drops below $1 M_{\odot}$ s$^{-1}$, and starts varying.

\begin{figure*}
\centering
\includegraphics[width=7cm]{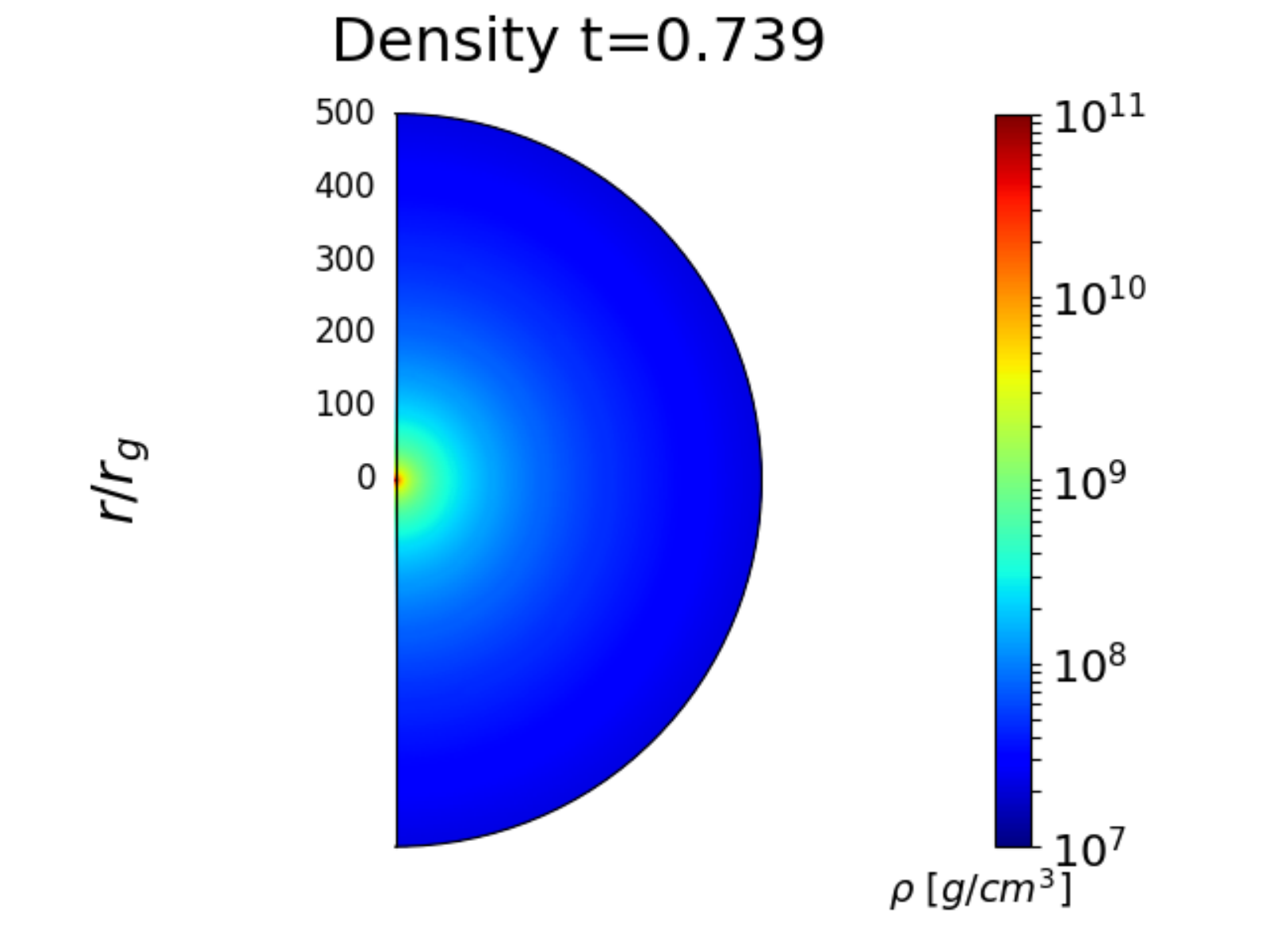}
\includegraphics[width=7cm]{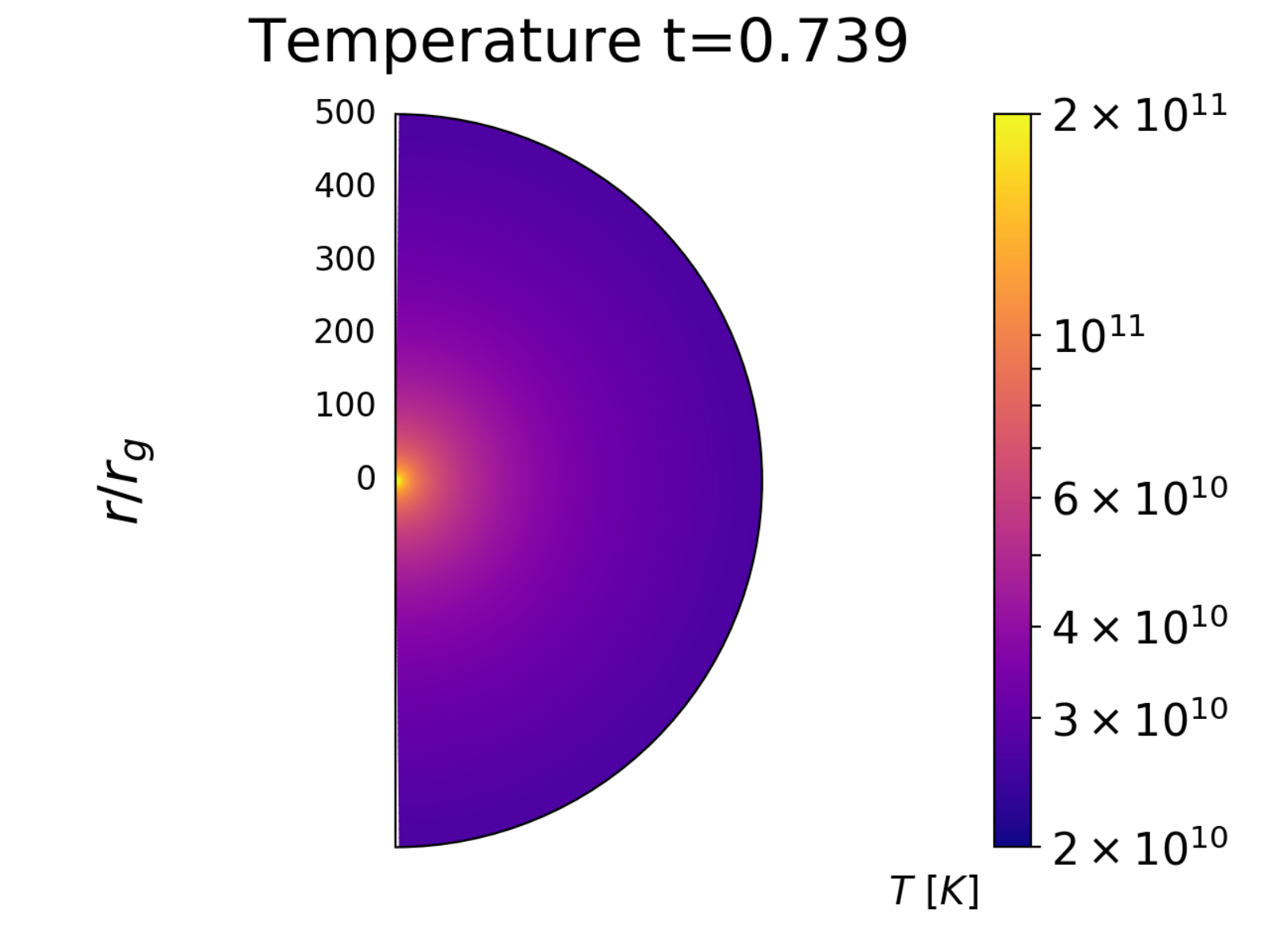}
\includegraphics[width=7cm]{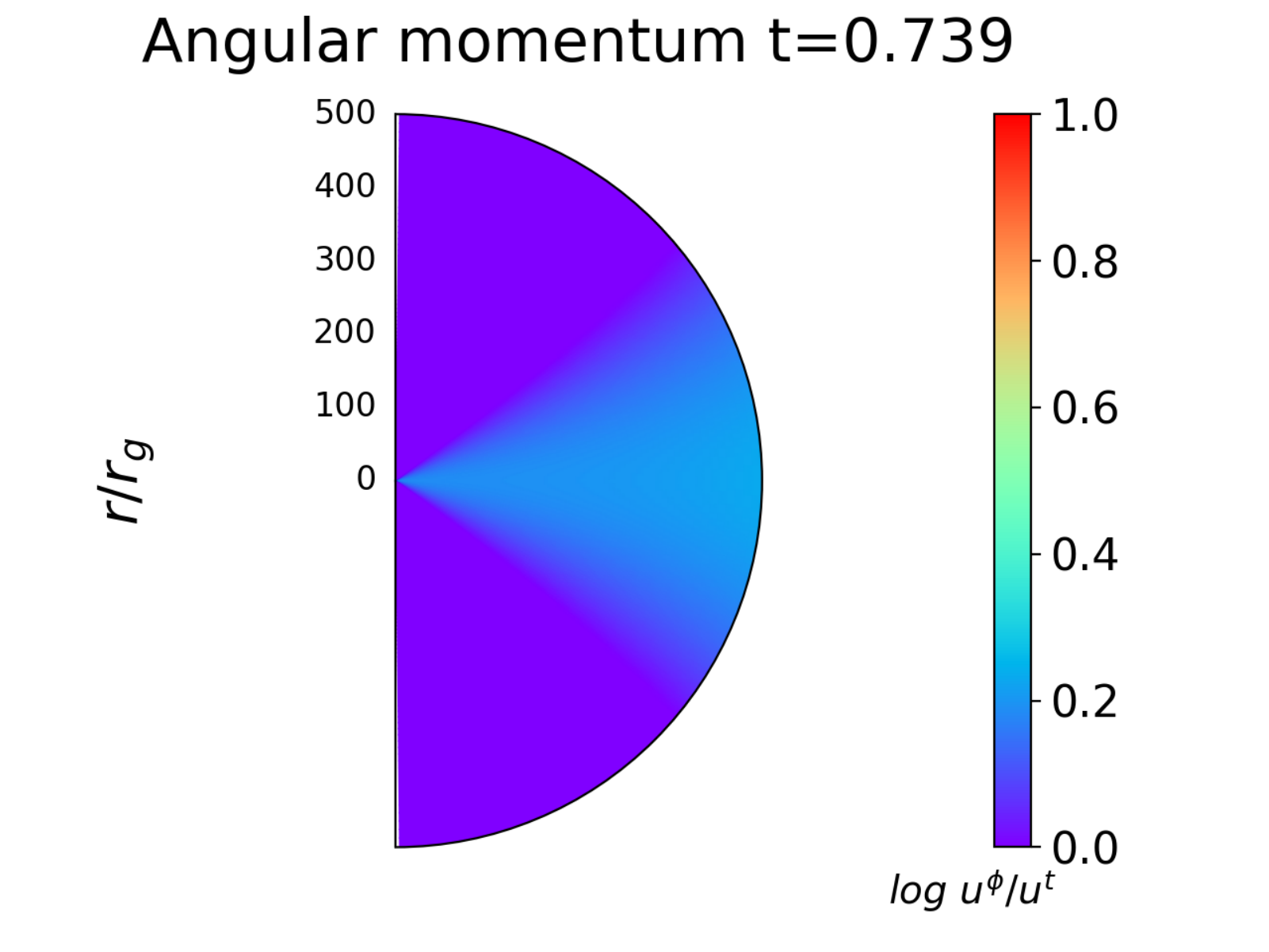}
\caption{Structure of the slowly rotating accretion flow,
at time t=50,000 M. The model assumes transonic flow with $r_{\rm s}=80 r_{\rm g}$, and
almost no rotation ($l/l_{\rm crit}=0.4$). 
 The maps show: (i) density,
  (ii) temperature of the plasma, (iii) specific angular momentum.
  Parameters: initial black hole mass $M_{\rm BH}^{0} = 3 M_{\odot}$, and initial
  mass in the cloud $M_{\rm cloud}= 25.0 M_{\odot}$. The maps show inner 500 $r_{\rm g}$.}
\label{fig:maps_rs80_l04}
\end{figure*}

\begin{figure*}
\centering
\includegraphics[width=7cm]{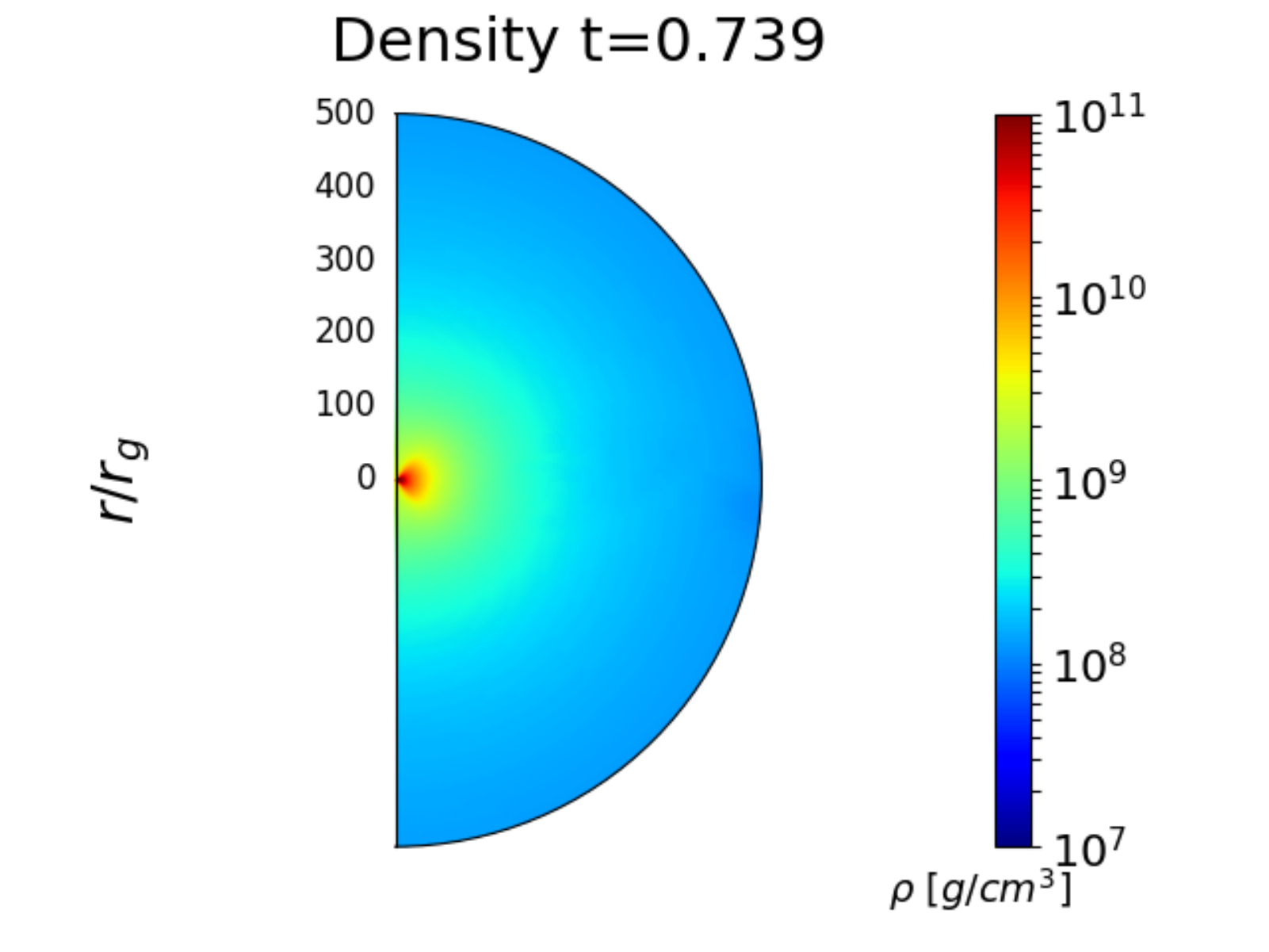}
\includegraphics[width=7cm]{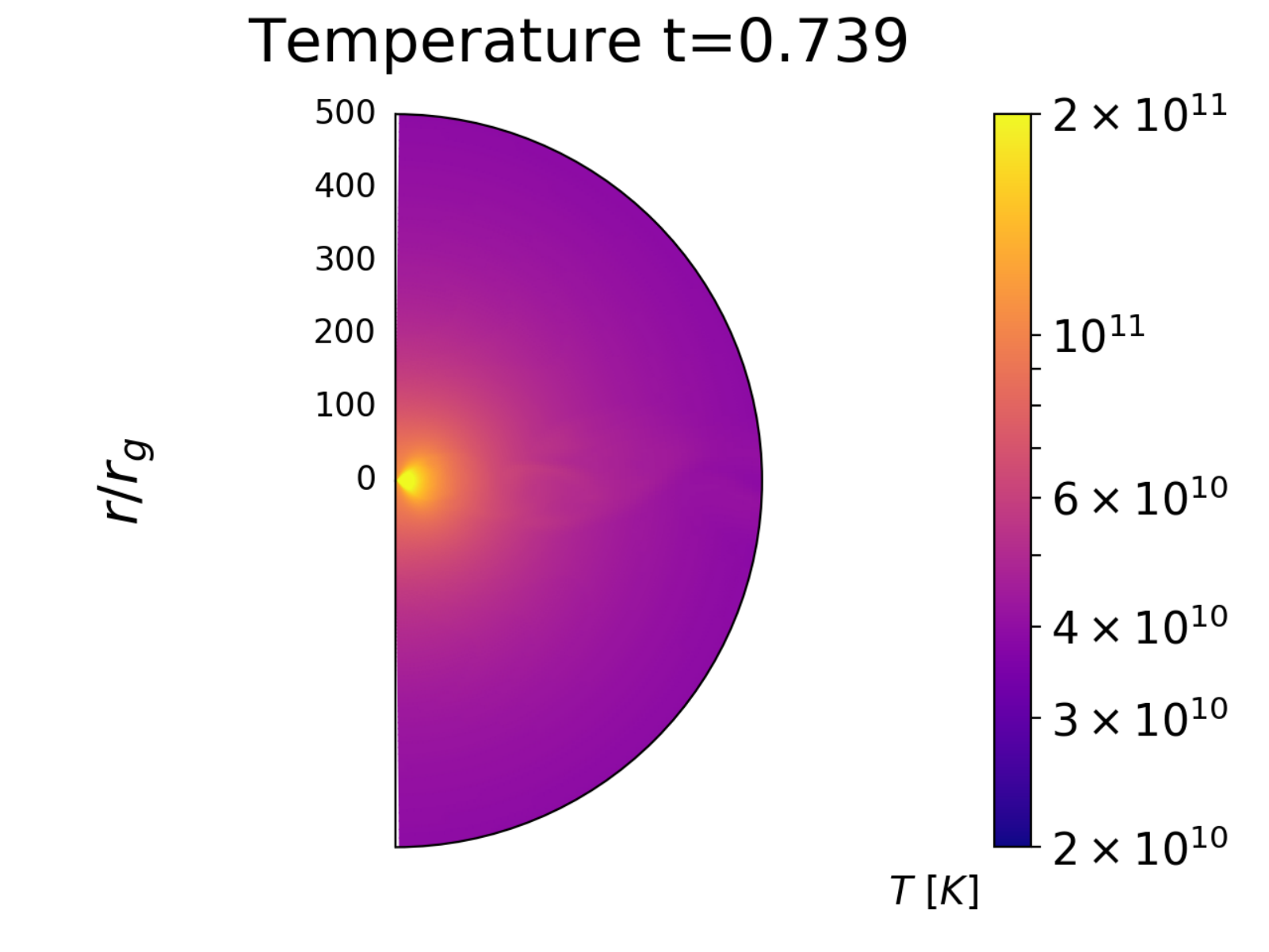}
\includegraphics[width=7cm]{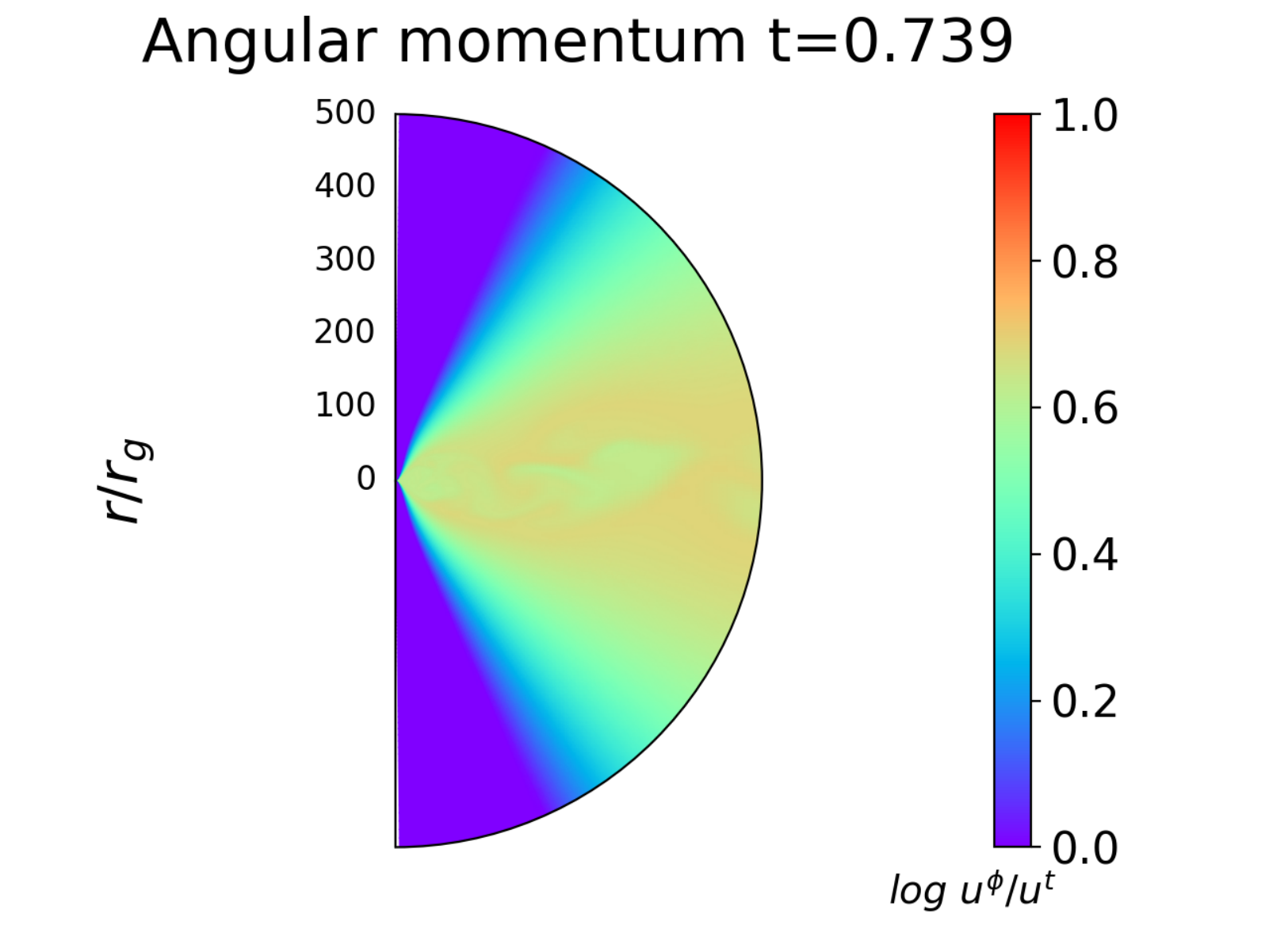}
\caption{Structure of the slowly rotating accretion flow,
at time t=50,000 M. The model assumes transonic flow with $r_{\rm s}=80 r_{\rm g}$, and slow rotation  ($l/l_{\rm crit}=1.4$).
 The maps show: (i) density,
  (ii) temperature of the plasma, (iii) specific angular momentum.
  Parameters: initial black hole mass $M_{\rm BH}^{0} = 3 M_{\odot}$, and initial
  mass in the cloud $M_{\rm cloud}= 25 M_{\odot}$. The maps show inner 500 $r_{\rm g}$. }
\label{fig:maps_rs80_l14}
\end{figure*}

In Figures~\ref{fig:maps_rs80_l04} and \ref{fig:maps_rs80_l14} 
we show the maps of the
flow structure calculated in the 2-D model, as taken in the inner 500 $r_{\rm g}$. The maps are plotted 
for two models of the collapsing star interior, 
where the transonic Bondi accretion flow is supplied with a small rotation.
The case of $l/l_{\rm crit}=0.4$ in fact presents almost no rotational support at the time of the snapshot, and the case of  $l/l_{\rm crit}=1.4$ presents a weak rotational support.
The snapshots, taken at the simulation
time t=0.739 s ($50,000 M$), present the rest mass density $\rho$ in the 
physical units, adopted to close the initial mass of the cloud in the domain at the 
level of $25 M_{\odot}$. The gas
temperature T is expressed in Kelvin. We also show the specific angular momentum distribution.

\begin{figure}
\centering
\includegraphics[width=7cm]{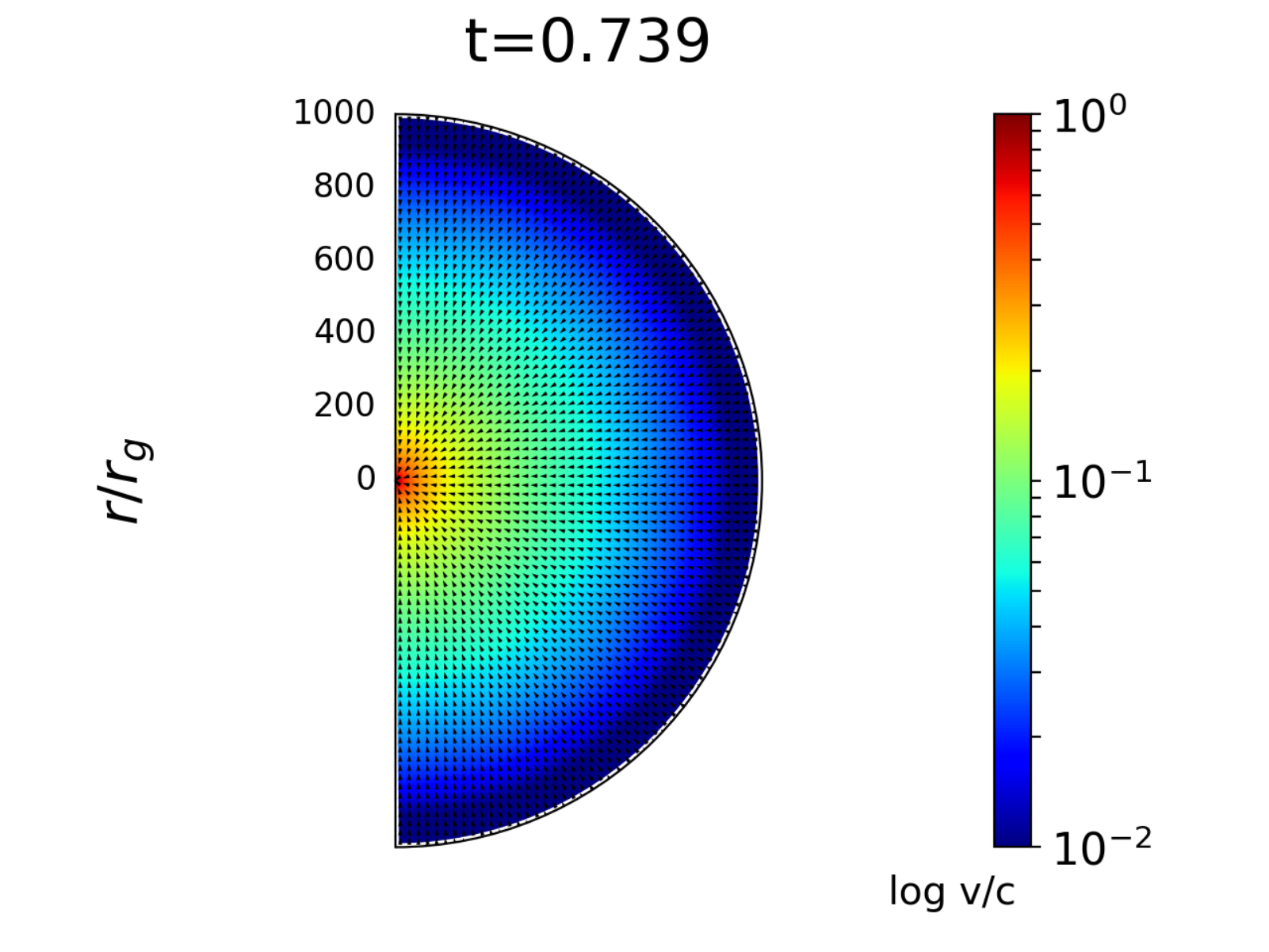}
\includegraphics[width=7cm]{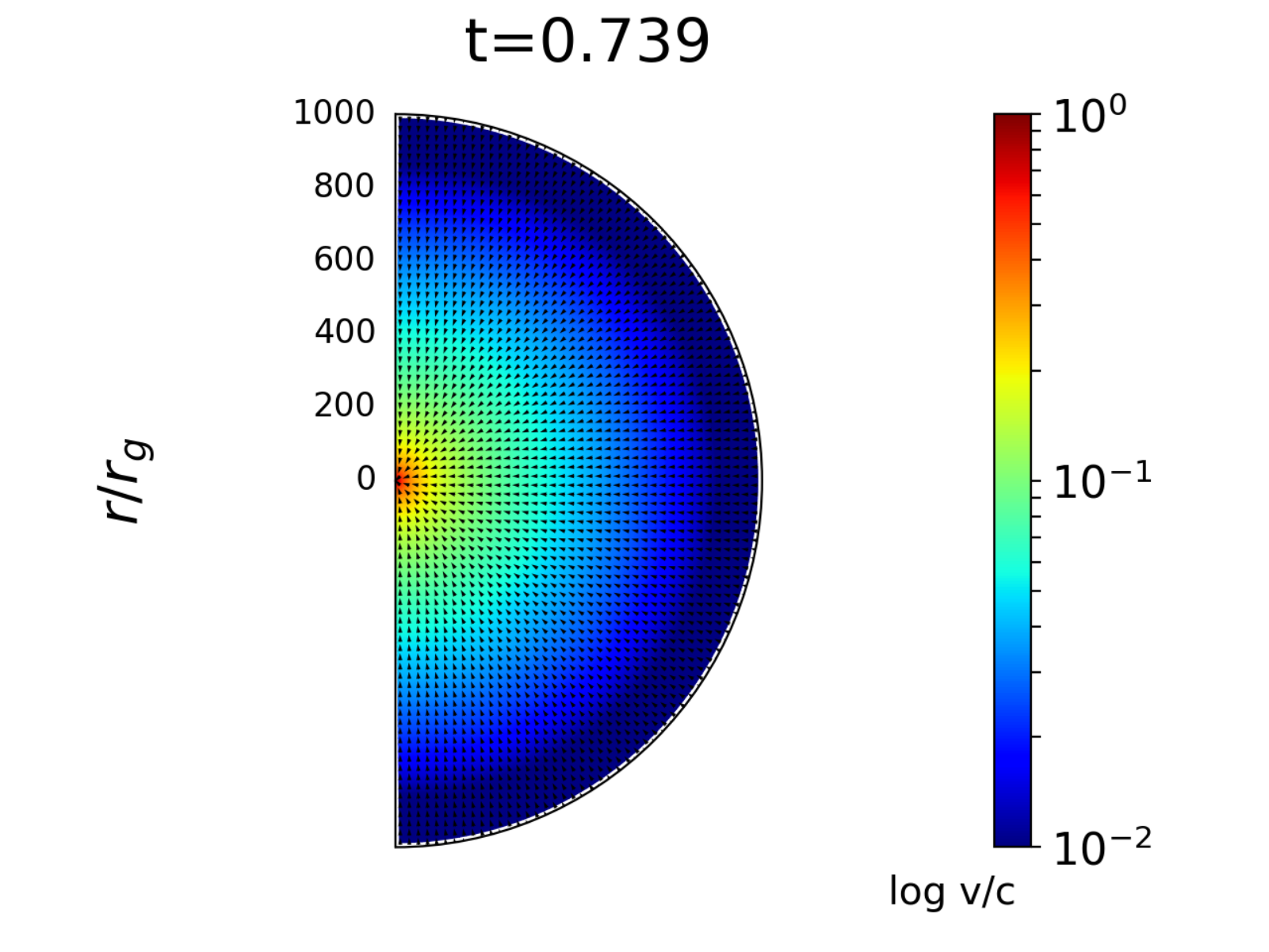}
\includegraphics[width=7cm]{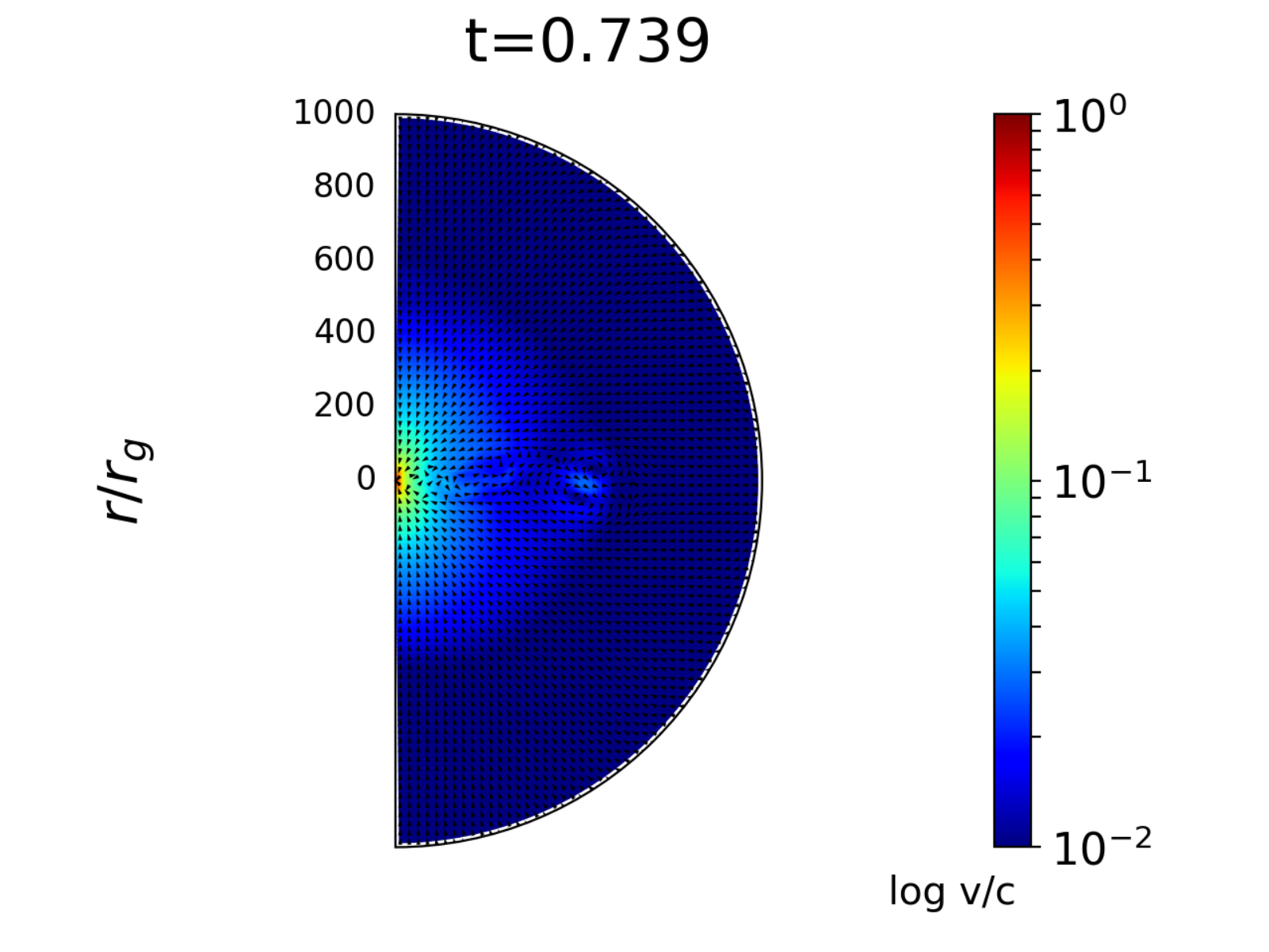}
\caption{Velocity fields at t=50,000 M, 
for the almost non-rotating ($l/l_{\rm crit}=0.4$, left), critical ($l/l_{\rm crit}=1.0$, middle), and weakly rotating ($l/l_{\rm crit}=1.4$, right) models. 
The cloud mass at this time is between $M_{\rm cloud}\approx 2.5$, and $20 M_{\odot}$, 
depending on the rotation.}
\label{fig:velfields}
\end{figure}

The velocity field maps at the t=0.739 s are shown in
Figure~\ref{fig:velfields}. They clearly show the presence of the mini-disk in the supercritical regime, when the non-radial components of the velocity
are developed. The flow is rather turbulent inside the disk, and 
gas with different amount of angular momenta is mixed within the disk body.
This manifests in the variability of the accretion rate at later times.

\section{Appearance of shocks in the flows}

\begin{figure*}
\centering
\includegraphics[width=7cm]{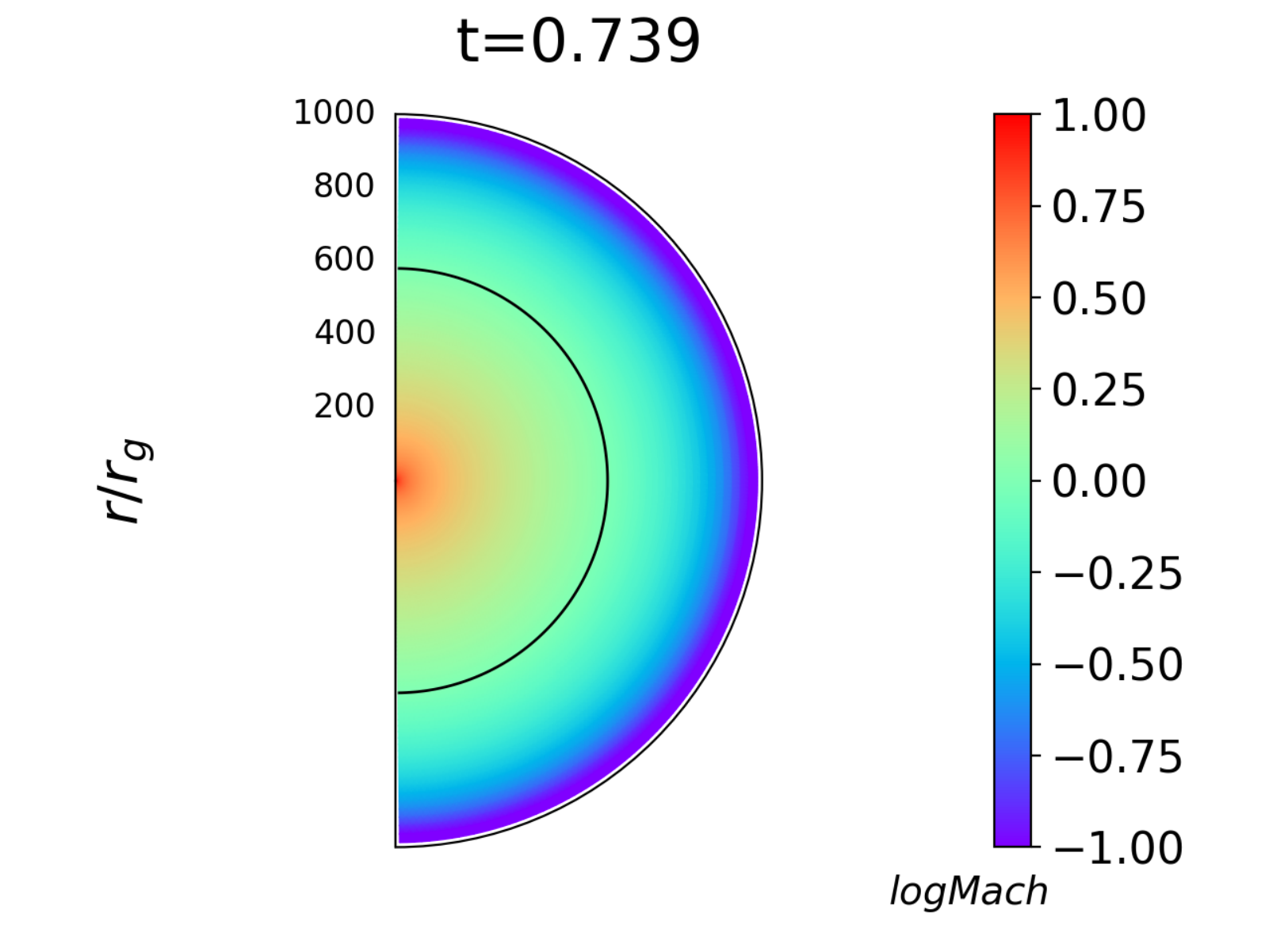}
\includegraphics[width=7cm]{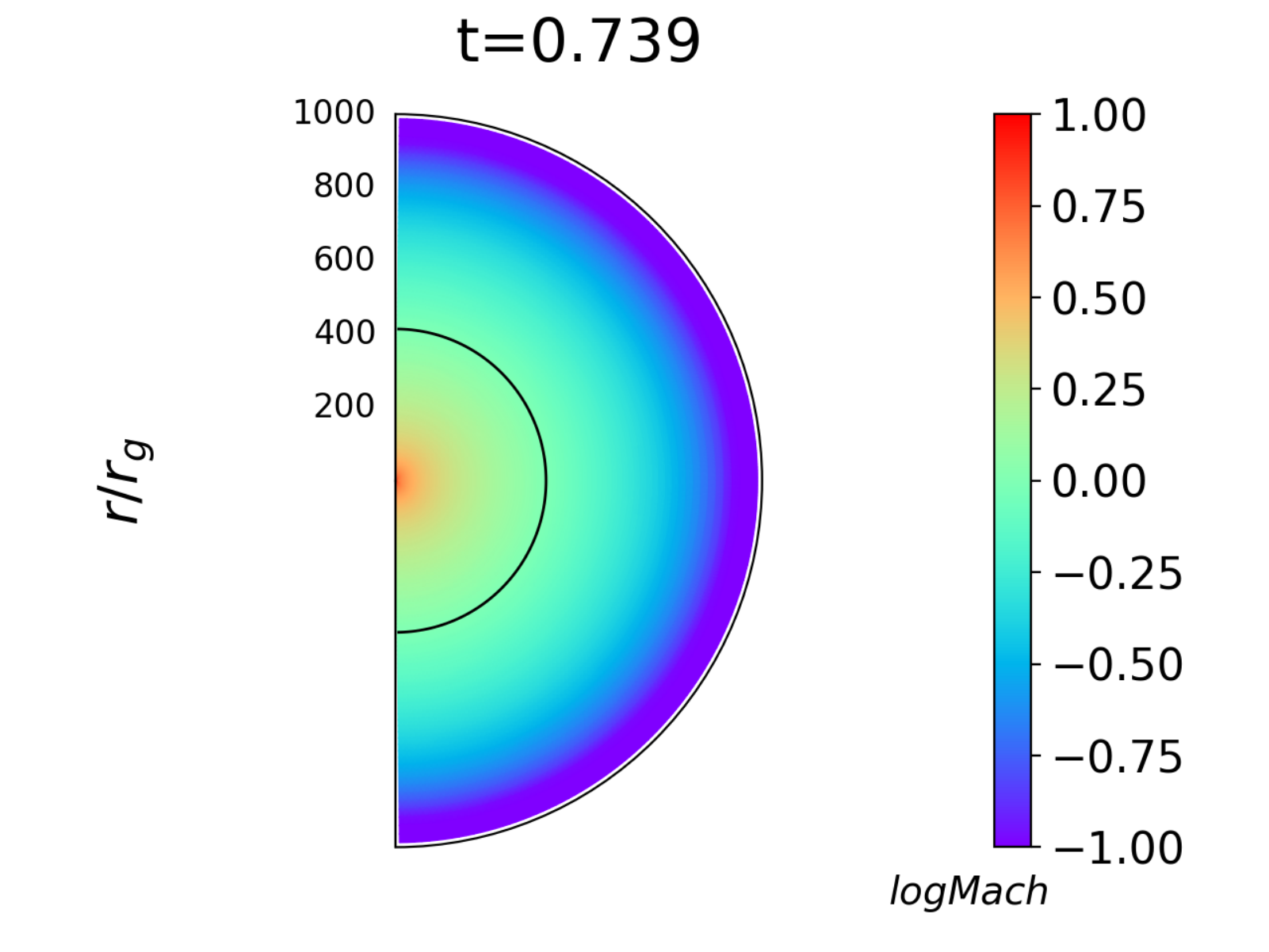}
\includegraphics[width=7cm]{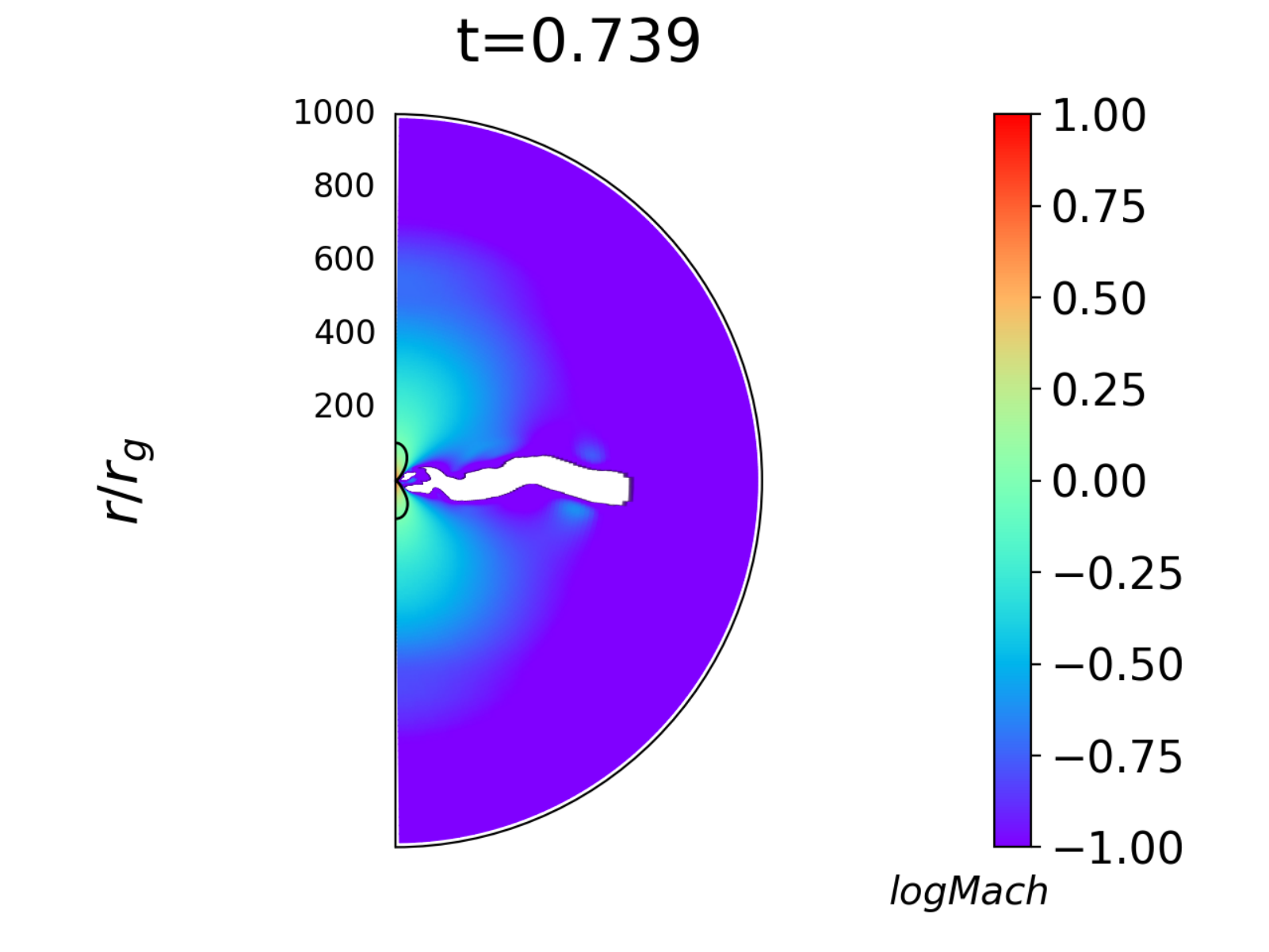}
\caption{Distribution of radial Mach number ($v_{\rm r}/c_{\rm s}$), 
at t=50,000 M, 
for the almost non-rotating ($l/l_{\rm crit}=0.4$, left), critical ($l/l_{\rm crit}=1.0$, middle)
 and weakly rotating ($l/l_{\rm crit}=1.4$, right) models. 
The thick solid line marks the sonic surface, i.e., the $M=1$ value.
The cloud mass at this time is $M_{\rm cloud} \approx 2.7 ~M_{\odot}$, or $7.4 ~M_{\odot}$, or $20~ M_{\odot}$, respectively.}
\label{fig:mach}
\end{figure*}

In Figure \ref{fig:mach} we show the profiles of the radial Mach number 
in the three simulations with different specific angular momenta, at the time $t= 0.44 s$ (50,000 M).
The sonic point, initially located at $r_{s}=80 r_{\rm g}$ in all the models, moves outwards and reaches about 500 $r_{\rm g}$ at this time in the simulation,  for the sub-critical rotation models, CL-04 and CL-10.
The sonic surface diffusion is slower in case of the critical model, CL-10,
and the sonic radius at this time of simulation is close to 300 $r_{\rm g}$.
For model CL-14, the sonic surface has no spherical shape. Most of the flow in the equatorial regions is
subsonic, and the Mach number in the 'mini-disk' is rather small.

\begin{figure*}
\centering
\includegraphics[width=7cm]{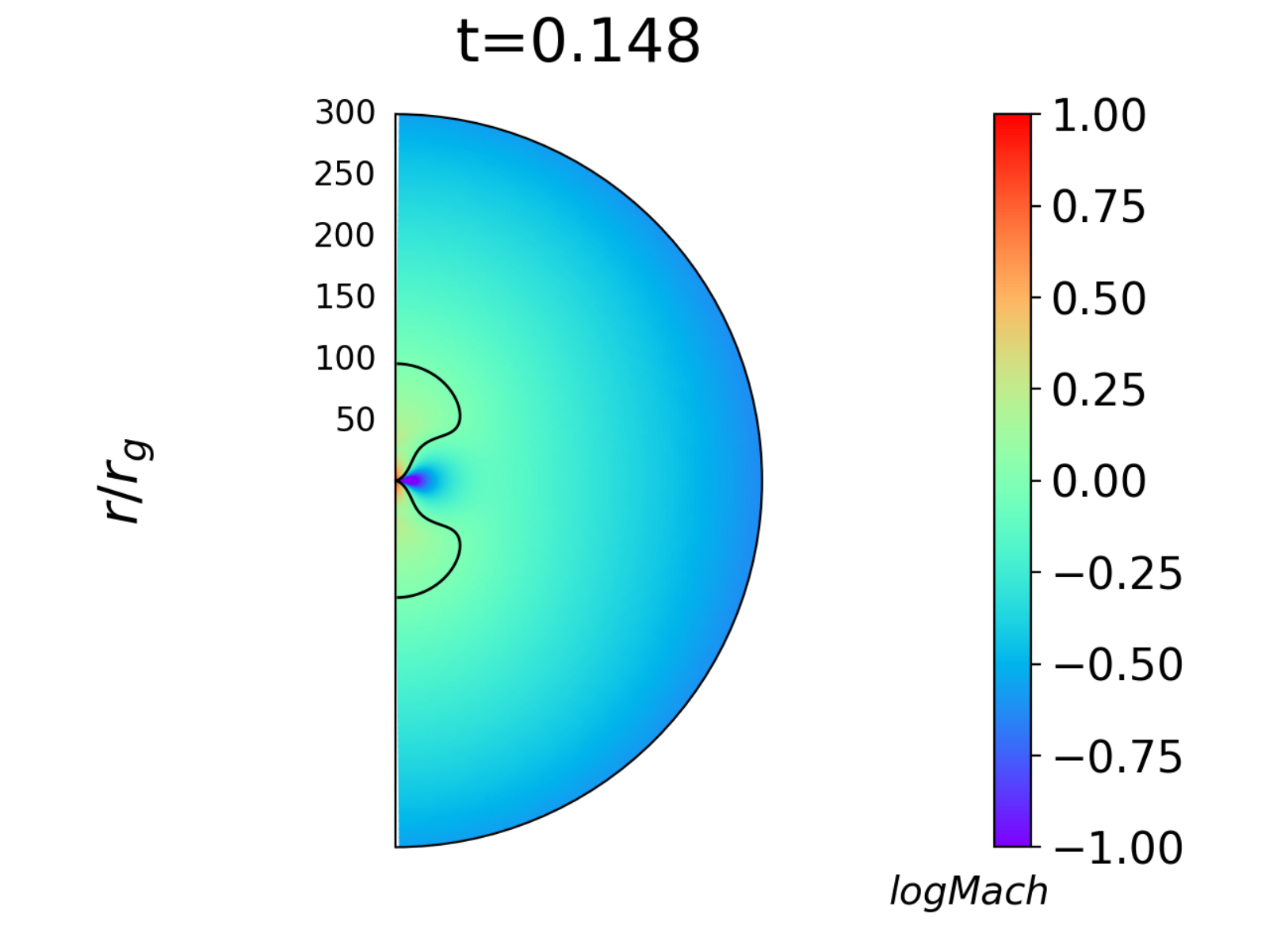}
\includegraphics[width=7cm]{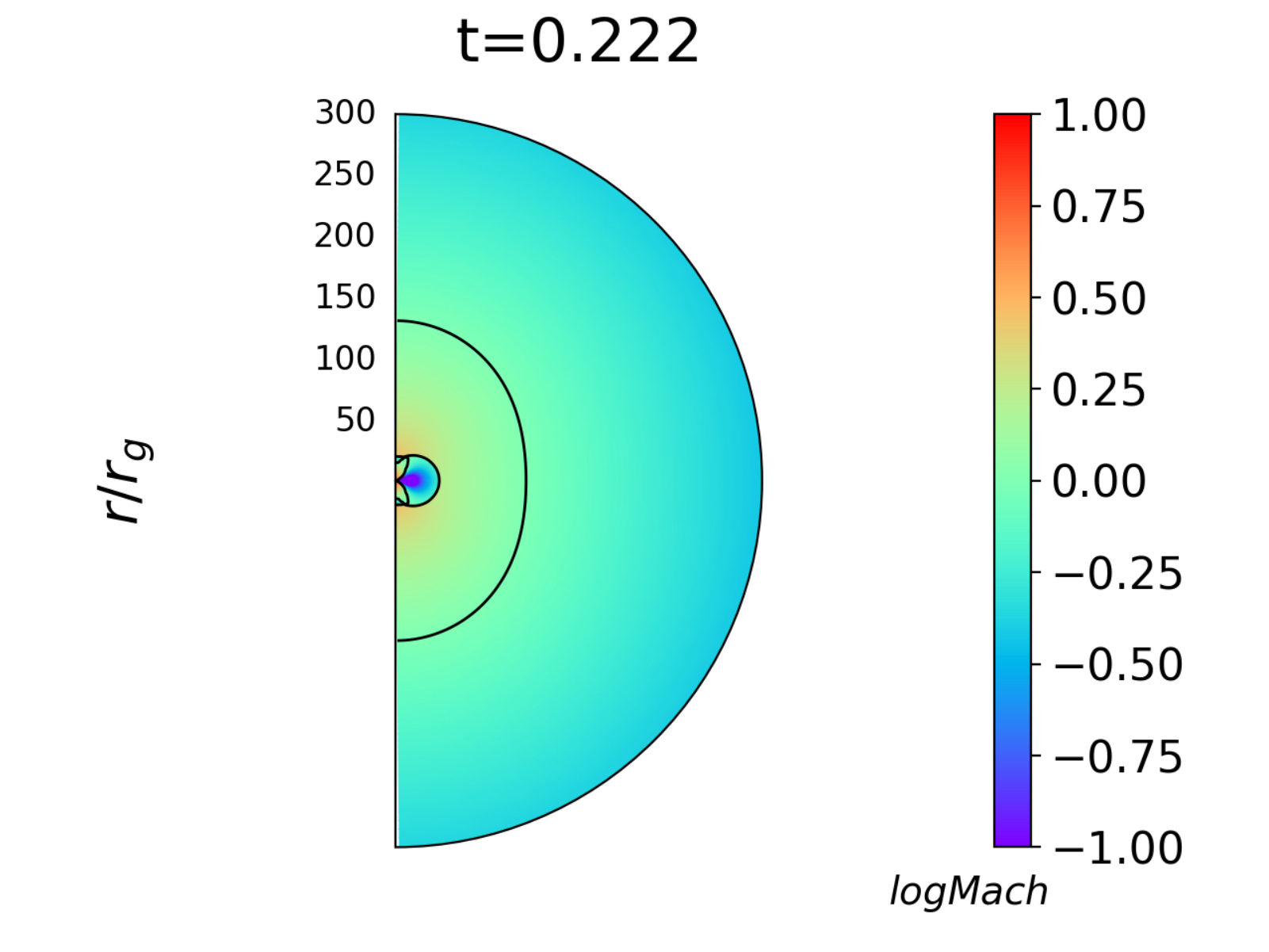}
\includegraphics[width=7cm]{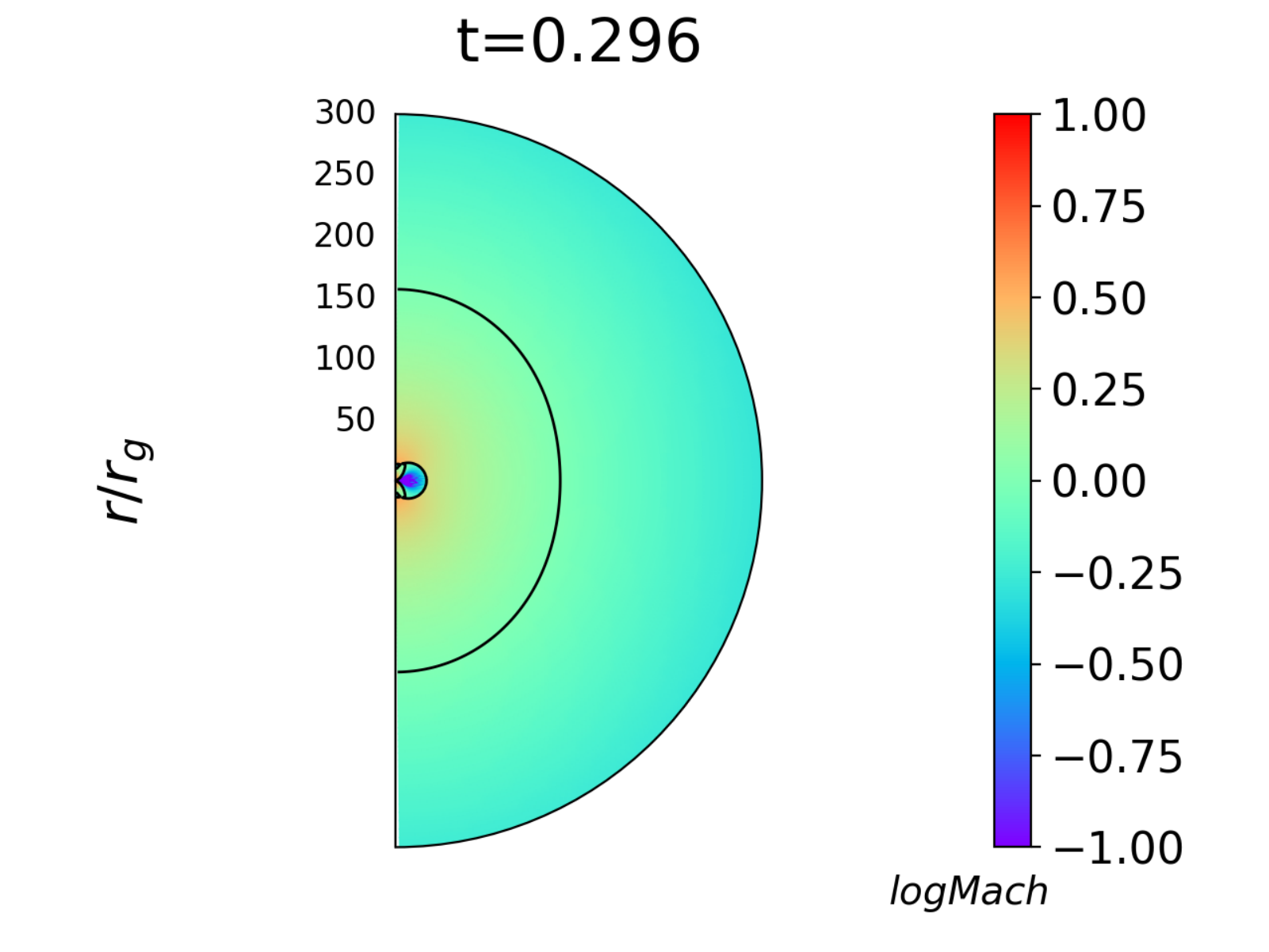}
\includegraphics[width=7cm]{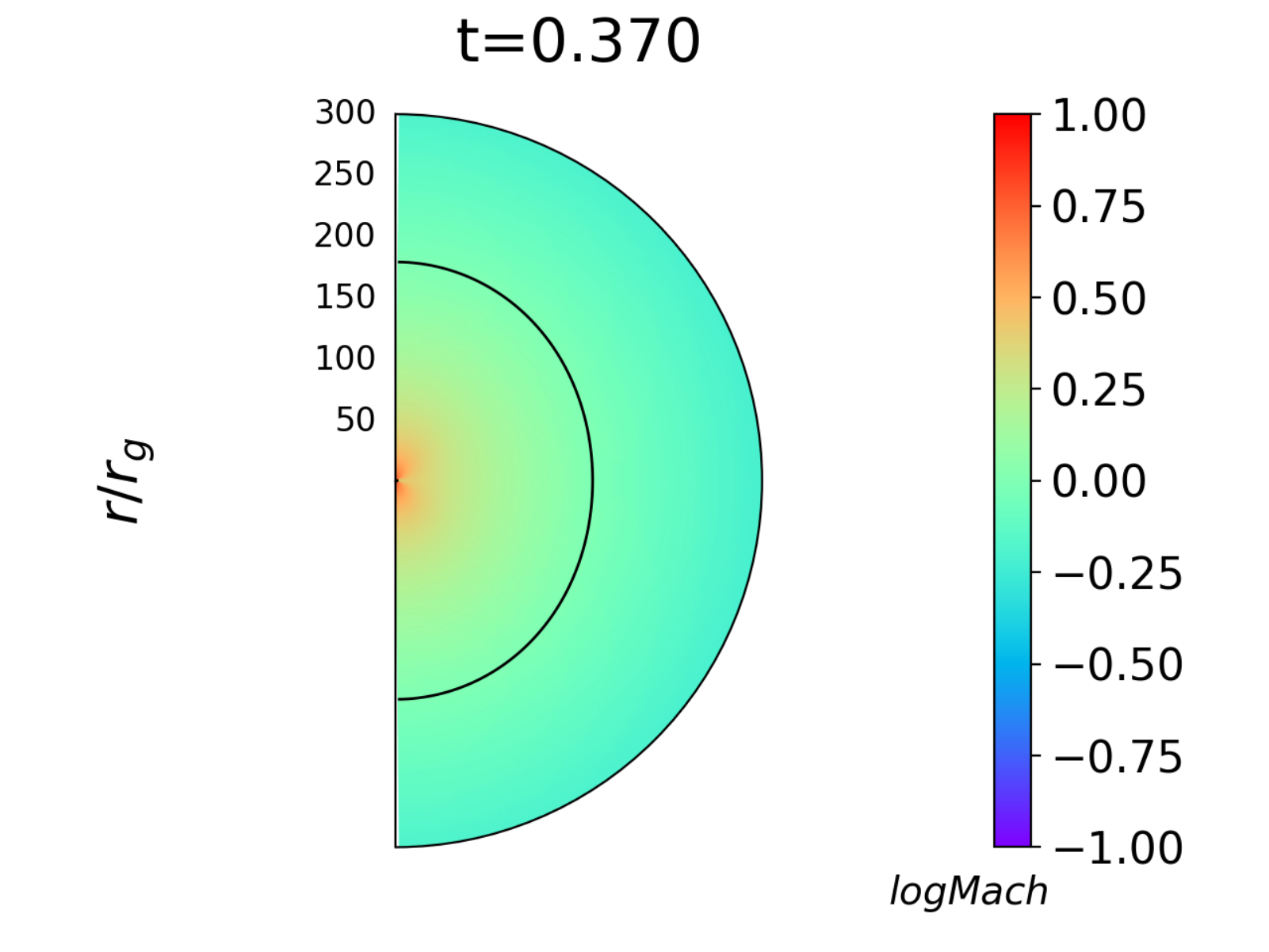}
\caption{Distribution of radial Mach number ($M=v_{r}/c_{s}$), 
for the critically ($l/l_{\rm crit}=1.0$)
 rotating model. 
The thick solid line marks the sonic surface, i.e., the $M=1$ value.
The snapshots were taken at times $t=0.148, 0.222, 0.296$, and 0.37 s, 
from left to right.}
\label{fig:mach_zoom}
\end{figure*}

In Figure \ref{fig:mach_zoom} we plot the zoomed-in distributions of the 
radial Mach number, and sonic surfaces, as taken for the critically-rotating model, CL-10.
The time snapshots were taken at the range of 0.14-0.37 s
(which corresponds to the geometric time
10,000, 15,000, 20,000, and 25,000 M).
As was seen in Fig. \ref{fig:accrate}, the accretion rate onto black hole
at this time is varying in this model.
We notice that this is the effect of the mini-shock bubble, that 
 appeared close to the inner radius, below $\sim$ 50 $r_{\rm g}$. 
The shock bubble was subsequently accreted onto black hole after that time.
The evolution of the shock front position is described in more detail below.

\begin{figure}
\centering
\includegraphics[width=0.45\textwidth]{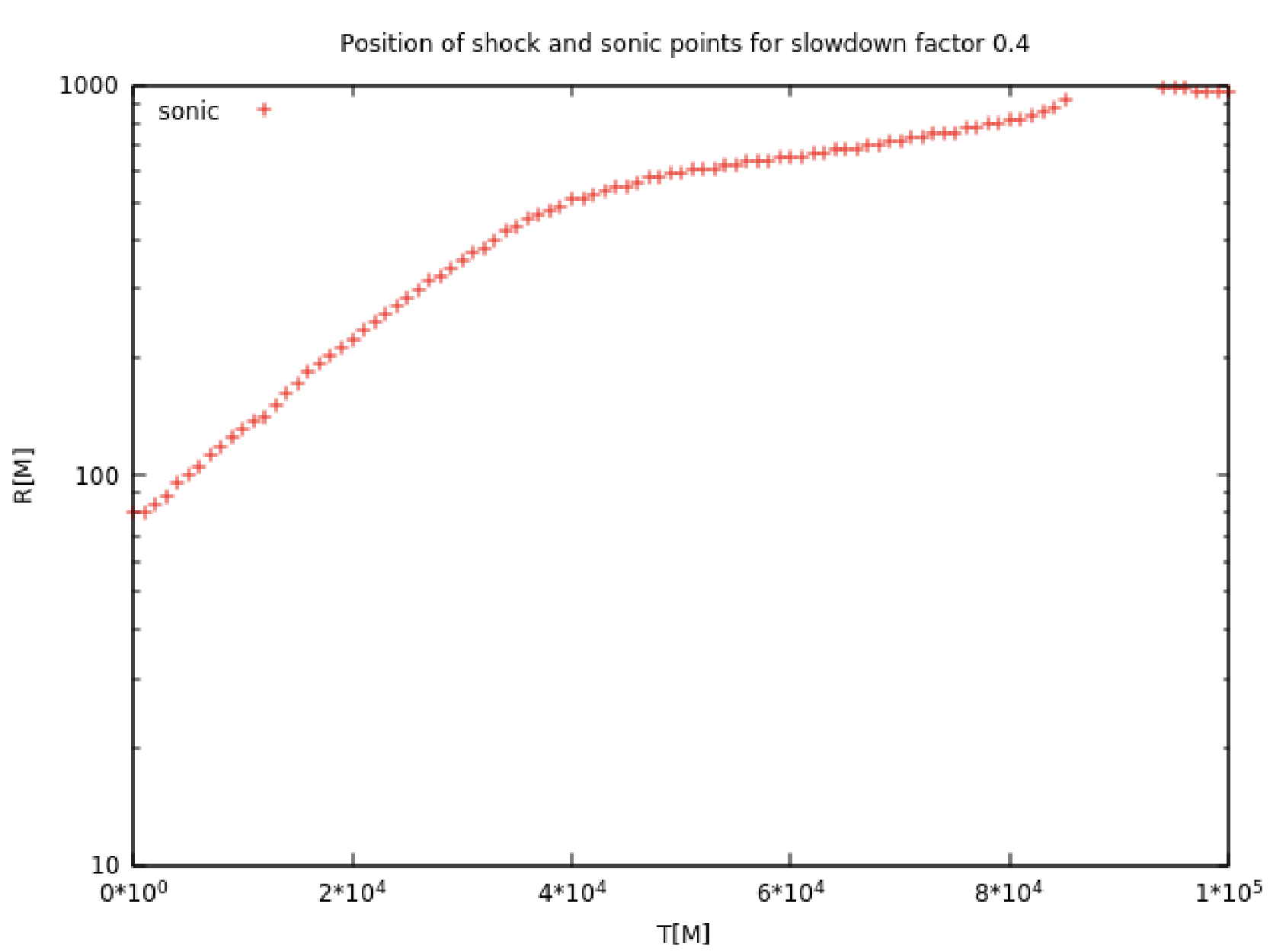}
\includegraphics[width=0.45\textwidth]{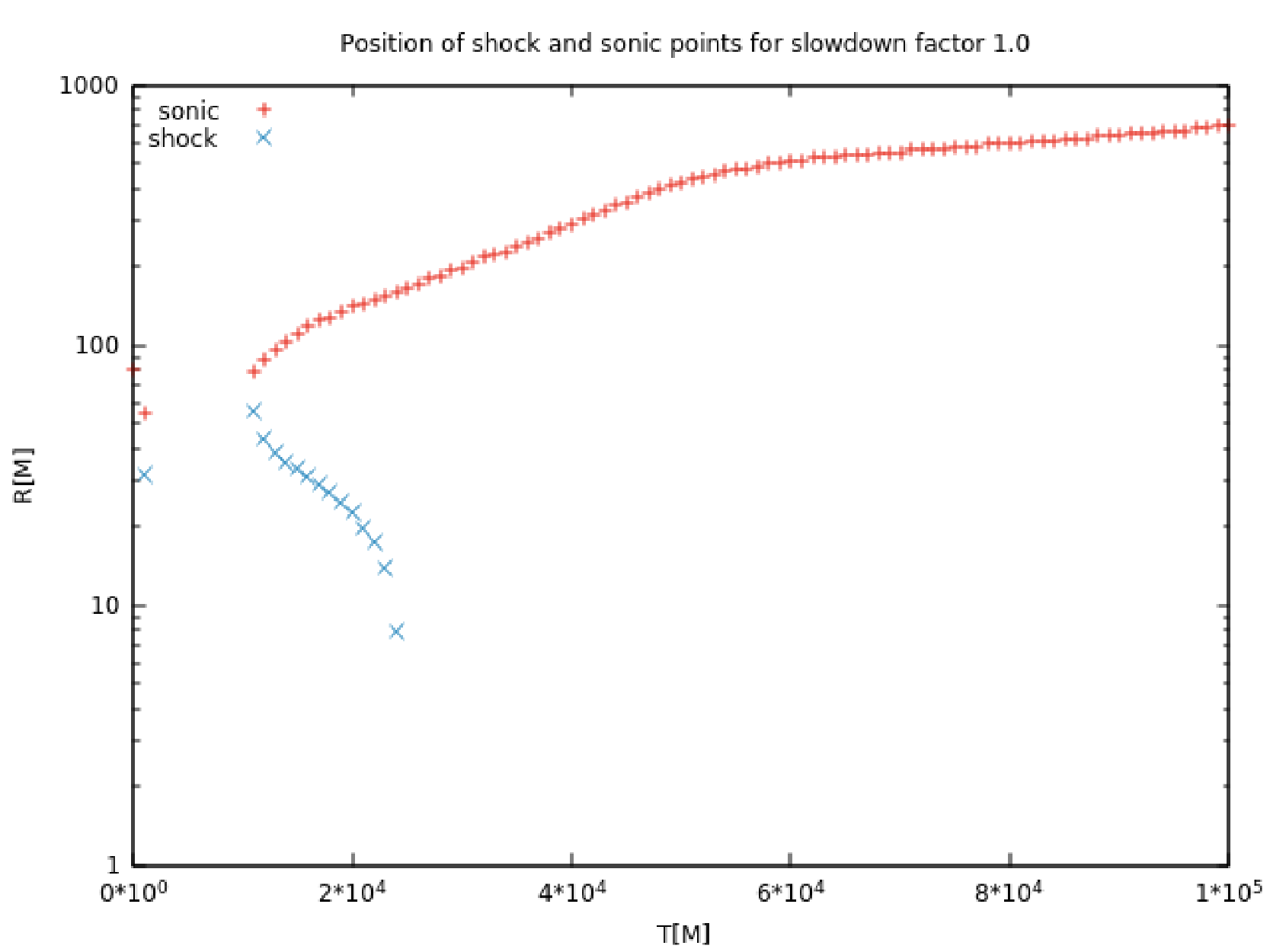}
\caption{
  Position of shock and sonic points in time, for the simulations with the
  initial 
sonic radius located at 80 $r_{\rm g}$ (marked with a red dot), 
and the slowdown factor of $S=0.4, 1.0$. The computational domain goes up to 
$R_{\rm out}=10^{3} r_{g}$ 
and the whole region is filled with transonic Bondi flow.
}
\label{fig:rshock1}
\end{figure}

In Figure \ref{fig:rshock1} we present the evolution of the shock and sonic points position with time, for models
SL-04 and SL-10.
All these results are plotted at the equatorial plane.
We note, that the supersonic parts of the flow for model SL-14 also exist, but
they are present in the polar regions, above and below the equator (cf. Figure \ref{fig:mach}).

The plots in Fig. \ref{fig:rshock1}
encompass the time of evolution for the long runs, up to $t=1.48$ s (100,000 M).
The cloud was initially filled with Bondi flow
and during the simulation time, the matter was supplied to the black hole from the outer regions. As indicated in Table \ref{table:models}, at the end of the long runs there was almost no gas left in the domain, unless it was kept on the orbit by rotation (run SL-14).

The shock position is computed from the profile of the radial Mach number, 
when the local sound speed is estimated as
\begin{equation}
c_{\rm s} = \sqrt{{4 \over 3} {P \over \rho + 4 P}}
\end{equation}
which holds for $\gamma=4/3$ adiabatic index.

The local radial velocity is computed as follows.
Under a change of coordinates the $ x^{\mu} \rightarrow  x^{\tilde{\mu}} $ components transform as
\begin{equation}
 \textsf{g}_{\tilde{\mu}\tilde{\nu}} = 
\frac{\partial \textsf{x}^{\rho}}{\partial \textsf{x}^{\sigma}} 
\frac{\partial \textsf{x}^{\tilde{\mu}}}{\partial \textsf{x}^{\tilde{\nu}}}
\textsf{g}_{\rho \sigma} 
\end{equation}
Therefore
\begin{equation}
(\begin{array}{ccc}
d\textsf{u} \\
d\textsf{v} \\
\end{array} ) = 
(\begin{array}{ccc}
\frac{\partial \textsf{u}}{\partial \acute{\textsf{u}}} &\frac{\partial \textsf{v}}{\partial \acute{\textsf{u}}} \\
\frac{\partial \textsf{u}}{\partial \acute{\textsf{v}}} &\frac{\partial \textsf{v}}{\partial \acute{\textsf{v}}} \\
\end{array} )
(\begin{array}{ccc}
d\acute{\textsf{u}} \\
d\acute{\textsf{v}} \\
\end{array} )
\end{equation}

We transform the velocity from the code coordinate $x^{\nu}(x^0,x^1,x^2,x^3)$ to  velocity in the Boyer-Linquidst coordinates, $ (t,r, \theta, \phi)$. 
The radial velocity in the code coordinates is $ V^r = \frac{dx^1}{dx^0} = \frac{d x^1}{dt} $, 
and the radial velocity in the Boyer-Linquidst coordinates is $ U^r_{bl} = \frac{dr}{dt} $.
Therefore, $ U^r_{bl} = \frac{dr}{dt} = \frac{\partial r}{\partial x^1} \frac{dx^1}{dt}$.

In each model one or more shock fronts develop during the evolution, which affect the accretion rate and subsequently the growth of the black hole mass and spin. 
The presence of the subsonic bubble ('mini-disk') changes the accretion rate along the equatorial plane, while the supersonic material accrets through the polar funnels. 
Because the shock bubble can develop vertical oscillation (see the results for the shock evolution in the case of constant mass and spin of the black hole summarized in \cite{sukova2017}), the presence of the shock front typically leads to the flickering of accretion rate with peaks reaching one order of magnitude.

For the subcritical case, SL-04, the shock develops only in the late stage of the simulation, at about $t=1.5$ s (100,000 M). At that time, the remaining mass of the cloud is very low, several orders of magnitude lower than the initial mass, hence the accretion rate is also very low. 
Therefore, it does not influence the final mass and spin of the black hole.

In the critical case SL-10 the rotation causes very fast appearance of the inner shock bubble, which is growing in the equatorial plane. The subsonic region merges with the sonic point very quickly, at about $t=0.03$ s (2,000 M). 
After that time accretion proceeds as subsonic flow.
However, as the mass of the black holes grows, material becomes supersonic close to the black hole, especially along the axis, and eventually it encompasses the subsonic gas on the equator, where the rotation is the fastest, forming another shock bubble at about $t=0.15$ s (11,000 M). 
This bubble halts part of the material before being accreted, so that the accretion rate slightly drops down temporarily $t=0.15-0.22$ s ($t \in (11,000-15,000)$ M).

The mass of black hole is however still increasing, mainly due to the accretion through the polar regions, and after a while it comes to the regime, where the shock bubble starts to oscillate.
The vertical oscillations of the bubble again cause the flickering of the accretion rate, however, now the flickering appears in the time, when the accretion rate is very high (see Figure \ref{fig:accrate} at $t \in (0.25;0.35)$ s). 
During next about 0.15 s ($10,000$ M) this shock bubble is accreted completely. 

Between $t= 0.35$ s and $t=1.5$ s (25,000-101,000 M), no shock is present in the flow, while the sonic surface is inflating up to about 700 $R_{\rm g}$. At $t=1.51$ s (102,000 M), yet another shock front appears, and this one is growing very fast, transferring the supersonic accretion in the subsonic one and causing more flickering. This happens, however, again in the very low accretion rate regime and it is not substantial for the black hole growth.

In the super-critical model SL-14, due to the rotation the shock forms and expands very fast, so that it merges with the sonic surfaces, leaving almost the whole accretion flow subsonic. Only in the very vicinity of the black hole, there is an inner sonic surface, which is not spherical anymore.
The locations of sonic points depend on the altitude, and characteristic shape of the sonic surface, close to the 'eight' letter shape, develops in the inner region of  the shock bubble, as it is inflating. The flow inside the mini-disk is fully subsonic, and the material becomes supersonic mainly along the black hole rotation axis.

In this case the accretion rate is substantially lower than in the two previous cases, so that the remaining mass in the cloud shrinks more slowly.
Hence,  the whole process takes longer time and is less energetic, leaving the black hole less massive, but more spinning.
The `eight' shape develops, when the shock expands. 
As the rotating gas approaches initial sonic surface, the rotation quickly modifies the surface shape.
The cause for its shape is the slowdown of the gas in the equatorial region due to centrifugal force.
In fact, the gas is pushed outward along the equator as the centrifugal force and gas pressure halt the inflow .

From comparison with Figure \ref{fig:mass_spin_evol} one can notice, that in case of CL-10 there is a jump in spin at the time around t=0.3 s. At that moment, the dimensionless spin is 0.99, and then the spin decreases again, to reach the stable value of about $s=0.65$ (see Table \ref{table:models}).
We also notice, that the jump in the spin for sub-critical rotation model
was not so high, and the stable value of shock front position is achieved when the constant black hole spin saturated late at the simulation,
at about $t=1$ s (70,000 M).

The sonic surface for RL models moves slower outwards, than for generic SL runs.
For instance, in the model with circularisation radius $r_{\rm c} = 10$,
and critical rotation (i.e., run RL-10), the sonic radius
at time $t=50,000 M$ (0.739 s) is located at about 200 $r_{\rm g}$, while
in the model $r_{\rm c} = 6$ (run SL-10) the sonic radius at the same time is at about 400 $r_{\rm g}$. The mass of the cloud was equal to about 15.5 $M_{\odot}$, and 7.3 $M_{\odot}$, respectively.
In case of supercritical rotation models, as noted above, the sonic surface is not spherically symmetric, however also its maximum size is much larger for
the run RL-14 at the end of the simulation, than for the run SL-14.

\section{Summary}
\label{sec:diss}

In this work, we calculated the general relativistic, hydrodynamical model
of collapse of the star's central parts and growth of the black hole due to accretion.
We calculated the changing Kerr metric coefficients,
due to the evolving mass and spin of BH, at every time step in the simulation.
Such scenario is relevant
for the new black hole formation and
a long GRB progenitor, provided that the resultant black hole spin is 
large enough to power the GRB jet via Blandford-Znajek process.
If not, the scenario is still
applicable
for the collapsar model without a powerful GRB emission. In this way 
it is used to put the limit for the maximum spin of the black hole, whose mass 
has grown substantially during the collapse.
Another constraint can be obtained for the maximum mass of the black hole, whose rotation is fast enough to power GRB emission.

We also provided here the constraints for the formation of a rotationally supported mini-disk in the center of collapsar. In this context, an important issue is the radiative feedback that may occur during the collapse of a star into black hole. It may easily deposit some additional energy in the stellar envelope and unbind the outer layers of the star \citep{batta2017}. This will halt accretion, resulting in a less massive black hole, even below the limit for a black hole mass obtained here (56\% of the collapsing envelope mass, obtained 
 for a super-critically rotating star). Our present model is still rather 
simplified in the sense that it neglects radiative processes, e.g., the neutrino transport, and therefore the feedback mechanism is not accounted for.
However, we may speculate that the most massive black holes detected by LIGO could have formed only if there was essentially no rotation in the progenitor star, hence all the envelope falls into the black hole radially and no feedback occurs. Such a black hole would also rotate with a spin of $s<0.2-0.3$ at maximum, hence no powerful GRB would be associated with this event.

These limits are imposed by the amount of angular momentum in the star's envelope. The masses of merging
black holes detected by LIGO gravitational wave signals are larger tan those typically found for stellar mass black holes in X-ray binaries, and have values betwee 10 and 30 $M_{\odot}$.
The results of our computations suggest that their spins should be rather small. The details depend on the rotational properties of the evolved massive stars which formed the LIGO holes.
As for today, no remnants of the past GRB activities were reported in the 
case of these black holes,
so the picture presented in our work seems consistent 
with black holes being formed by direct collapse of a very slowly rotating
stellar core.

In our calculations, the central 
engine model of a GRB takes
 into account the non-stationary spacetime, 
due to the evolution of the black hole parameters. Previously, our computations were conducted in a fixed background metric in the context of a short GRB central engine modeled with high-angular momentum torus accretion (see \citet{janiuketal2013, janiuk2017}). The computations were based on the code, that utilized stationary geometry of spacetime, as is relevant for sub-Eddington accretion rates in 
active galactic nuclei, rather than GRBs. 
The present paper goes beyond such simplification.
We have shown here that accounting for the metric change affects quantitatively
the results for the speed of the accreting cloud evacuation. In particular, neglecting the black hole growth
led to some overestimation of the mass loss from the system through the outer boundary, i.e. winds.

Previously, in the work by \citet{sukova2017}, we also showed different regimes of accretion of low angular momentum flows in case of constant background Kerr metric. The existence of multi-sonic soulutions is
determined by the interplay of mass and spin of the black hole and angular momentum of the gas.
Here we study, how the feeding of the black hole by accretion changes this picture.
Because the black hole parameters are changing, while the angular momentum of the gas does not, the mutual relation of the parameters also changes. Thus different regimes of accretion and shock behavior are emanating during the evolution of the system. 

The presence of shock fronts in the gas on the other hand influences the growth of the black hole and its spinning up. 
If the combination of the three above mentioned parameters leads to the formation of the shock, less material accretes along the equator and the accretion rate could be temporarily lowered. 
If the shock bubble oscillates, flickering of the accretion rate occurs.
Consequently the time evolution of the black hole parameters is nonlinear.
Depending on the rotation profile of the star, even when the total mass of the cloud (collapsar) is the same, the final values of black hole parameters are affected by shock behaviour.

Because the process of accretion is believed to power the outgoing jets from the collapsar, the flickering and other changes in accretion rate translate to the jet propagation, thus leading to observable consequences. 
If the Lorentz factor of the outgoing blobs in the jets depends on the time varying accretion rate, then shells with different speeds can be produced. 
Their collisions far away from the central engine can produce highly energetic particles and radiation \citep{piran2004}.
In this way the shock presence can be important for the details of the lightcurve and energy spectrum of the measured GRBs.
Most important is the influence of a shock on the total duration of the event, and question whether that shock
is able to halt accretion.
The super-critical regime of rotation shown in our work proves that maximally spinnig black hole
can be supported by the mini-disk rotation as long as there is enough material in the collapsar
envelope to be accreted through it. The accretion rate at late times in this model is non-zero, but it is
highly variable.
Here, however, the feedback mechanism mentioned above may
quantitatively change the picture.
The study of this mechanism is the subject of our future work.

We note that in this work we explored only a limited
parameter space and the presented plots cover only
their fiducial values.  Our goal was to introduce the method, which is
explored here for the first time in application to the collapsar model,
and to briefly highlight one of its possible outcomes.
The initial mass of the black hole of $3 M_{\odot}$ is however
quite representative value, and agrees well with the results
of the core-collapse simulations (cf. $M_{BH}=2.6 M_{\odot}$, see \citet{kuroda2018}).
As for the mass of the progenitor star, the numbers used in various works range from above 70 Solar masses \citep{woosley2002},
down to 15 Solar mass \citep{lentz2015}. 
The 70 Solar mass pre-collapse star of \citet{takahashi2014}
in fact had the mass enclosed up to the Helium layer equal to the 31 Solar mass.
Our choice of the 25 Solar mass being enclosed in our computational domain of the size of
1000 gravitational radii meets the above constraints.
For the initial spin of the core, we presented here the simplest case of a zero spin.
We also checked, that other values (such as $a_{0}=0.3$ or $a_{0}=0.6$), lead to a very similar qualitative behaviour
in the simulation, and the super-critical rotation case is the only one which results in the maximum final spin of the black hole.
In this case, the black hole mass growth 
is only modest. The sub-critical rotation models, on the other hand, result in more massive black holes, which are not spinning fast. In fact, the dimensionless spin value at the end of the simulation being smaller than the initial one is also possible, if only the envelope was not sufficiently fast rotating.

We conclude therefore that the qualitative behaviour of the flow does not depend much on these parameters, and is a generic feature resulting from our approach.
The main uncertainty lies in the assumed rotation profile of the collapsar. This profile is however the main unknown, which was not solved yet neither by the stellar evolution models, nor by the massive stars observations.

\section*{Acknowledgments}

We thank Kostas Sapountzis for many helpful discussions.
This research was supported in part by grants DEC-2012/05/E/ST9/03914
and DEC-2016/23/B/ST9/03114
from the Polish National Science Center.
The simulations were performed on the supercomputer cluster of the 
Interdisciplinary Center for Mathematical Modeling of the Warsaw University, 
under computational grant GB 70-4.
PS is supported from Grant No. GACR-17-06962Y.

\appendix

The analytical solutions for the low-angular momentum
black hole accretion in the
evolving space-time metric do not exist.
However, in order to verify the performance of our code and test the
method on the simplest examples, we performed two types of additional
computations, in which the
numerical solutions can be compared with those for the stationary Bondi case.
The first type of test assumes zero, or very low, angular momentum
accretion onto the black hole, when the mass of the accreting cloud is
negligibly small in comparison with the mass of the black hole.
This computation we call here the 'Light Cloud' test.
The second test assumes a heavy cloud, which mass ratio to the
initial black hole mass is the same as in the main text,
but now we neglect the angular momentum. This computation is called `Pure Spherical' test.

\section{Light Cloud test}

\begin{figure}
  \includegraphics[width=7cm]{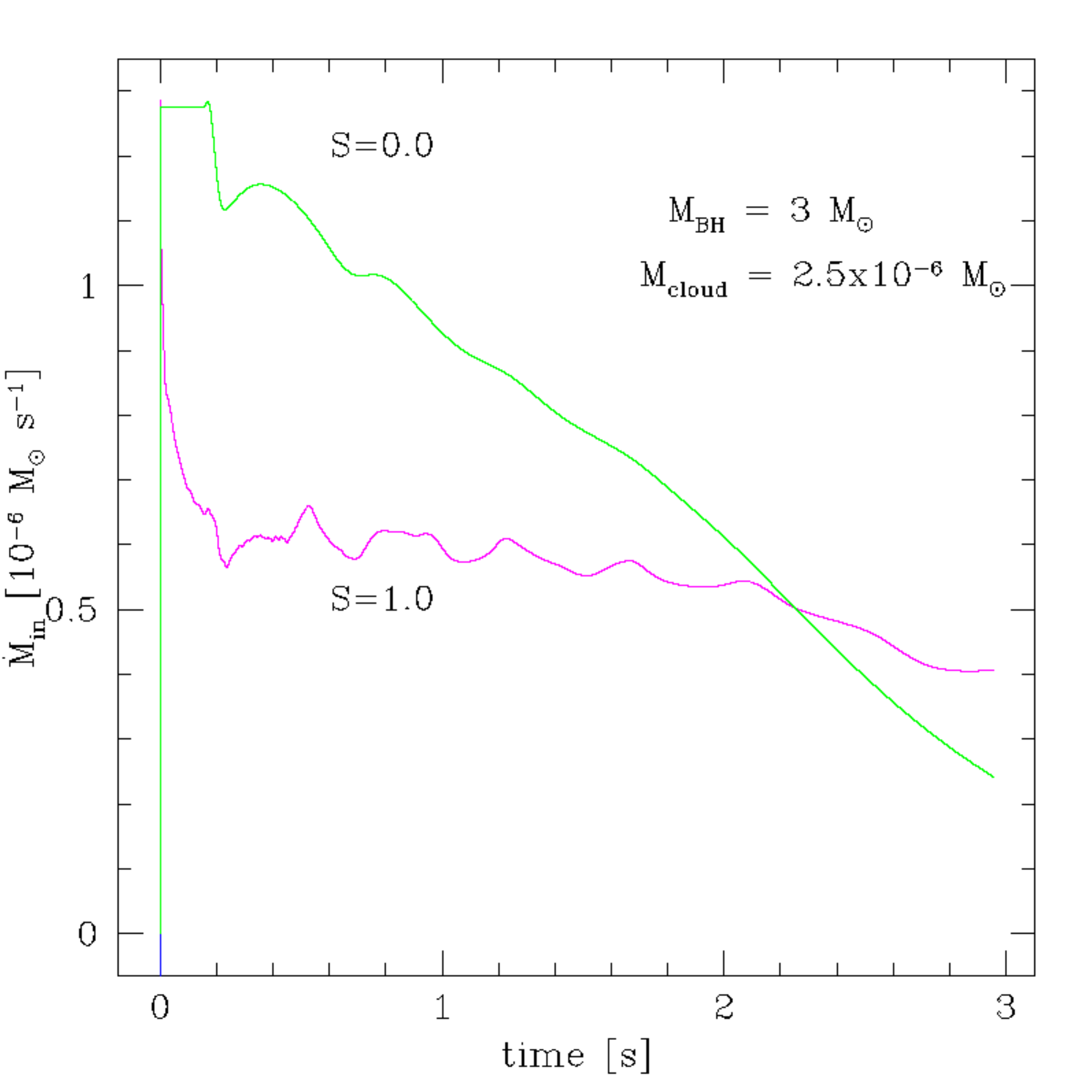}
  \includegraphics[width=7cm]{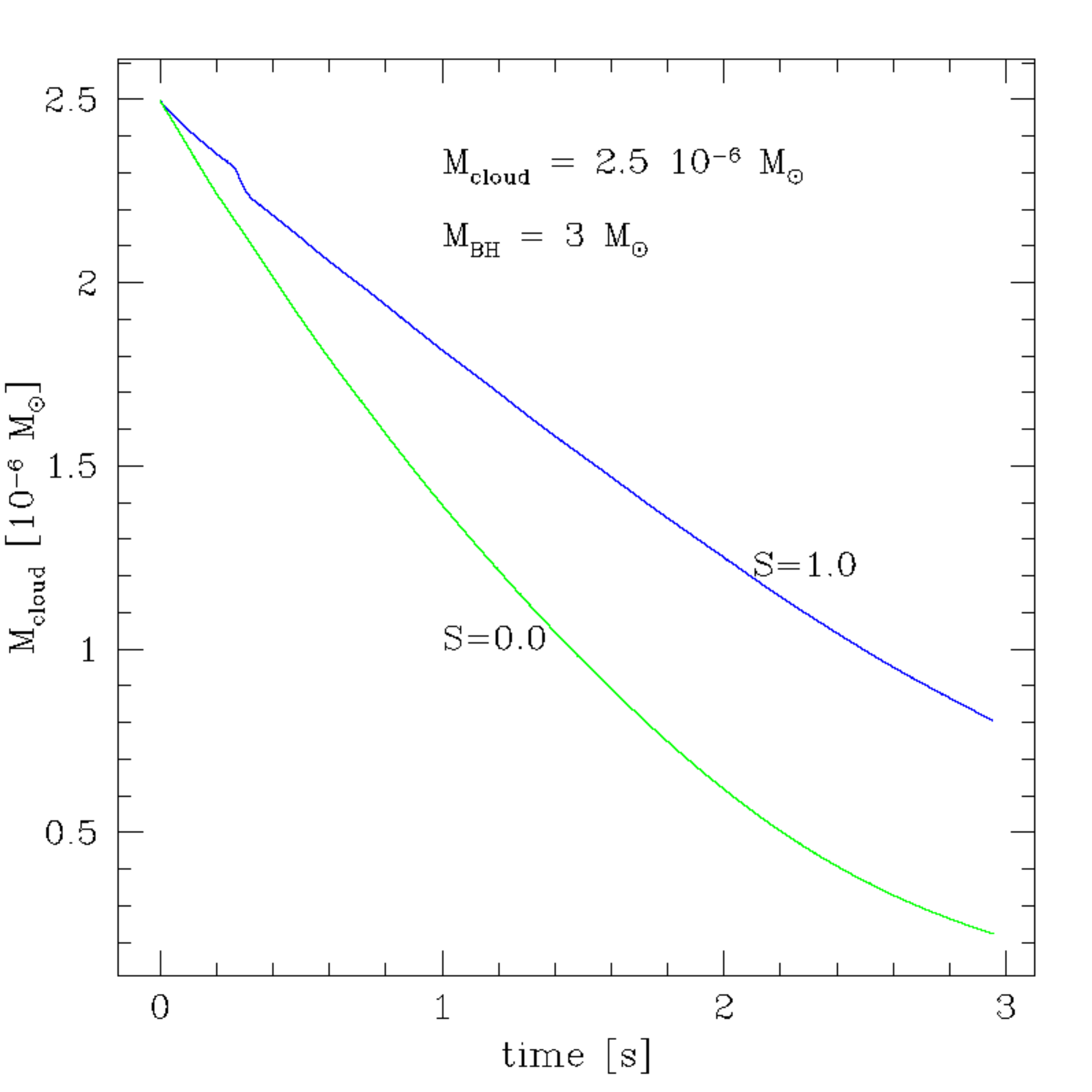}
  \caption{Mass accretion rate as a function of time, for a very light
    cloud accreting onto a $3 M_{\odot}$ mass black hole. Green line marks the solution without rotation, and blue line marks the solution with a slow rotation. The static and changing metric solutions are exactly the same. Right panel shows the corresponding total, volume integrated, mass of the cloud as it decreases in time.}
  \label{fig:mdot_light}
\end{figure}

The first test assumes that the mass of the cloud, which is surrounding the $3 M_{\odot}$ mass black hole, and contained within the volume of the size
$R_{\rm out}=1000$ gravitational radii, is equal to $2.5x10^{-6} M_{\odot}$.
As shown in the Figure \ref{fig:mdot_light}, the mass accretion rate
is very small, so that the mass of the black hole does not grow during the simulation. Neither changes the BH spin. We checked that it stays equal to $a=0$ athroughout the simulation, and in consequence the metric update
terms are practically not affected. Therefore the solutions do not depend
on the metric update routine, even though this part of the code is activated.

\begin{figure}
  \includegraphics[width=7cm]{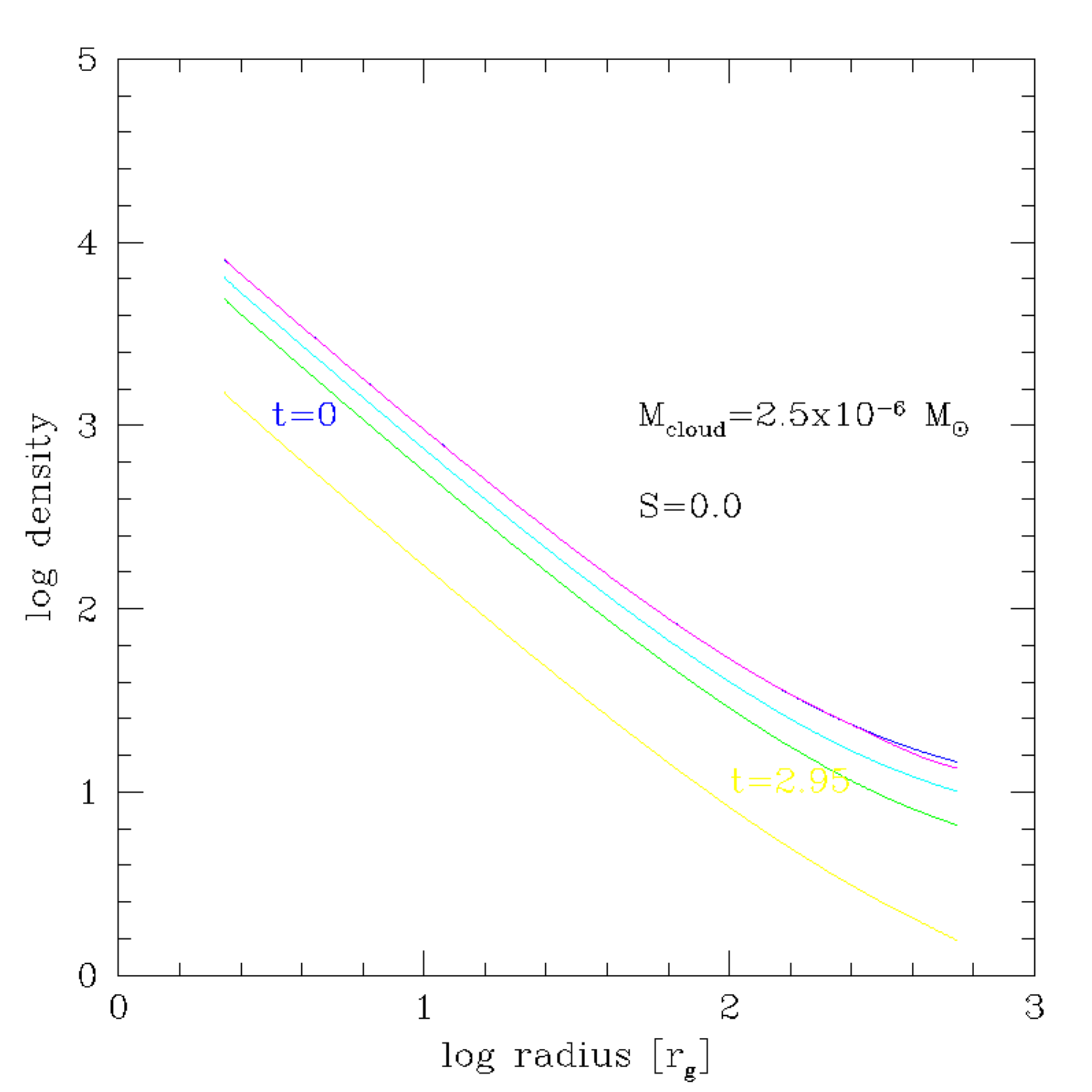}
  \includegraphics[width=7cm]{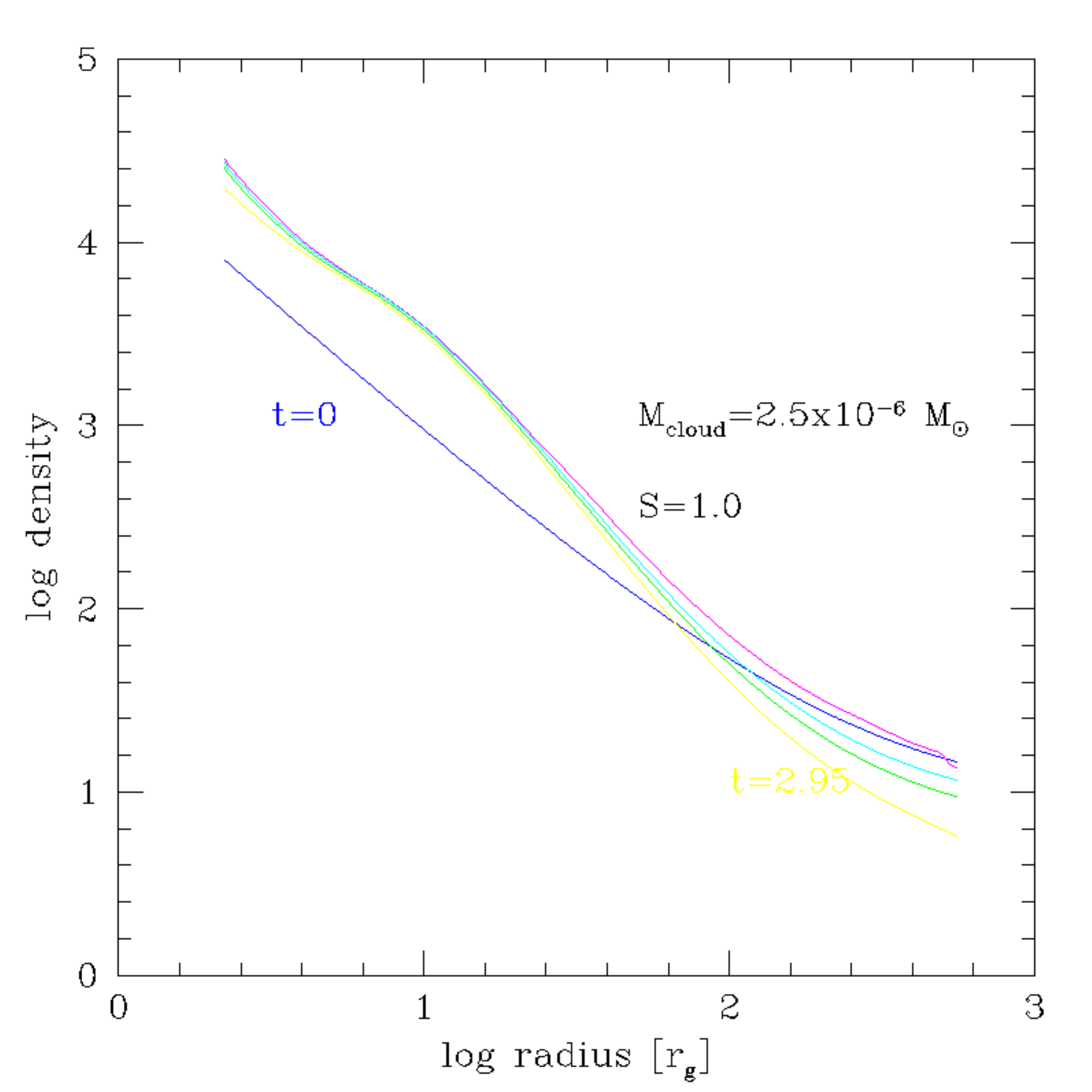}
  \includegraphics[width=7cm]{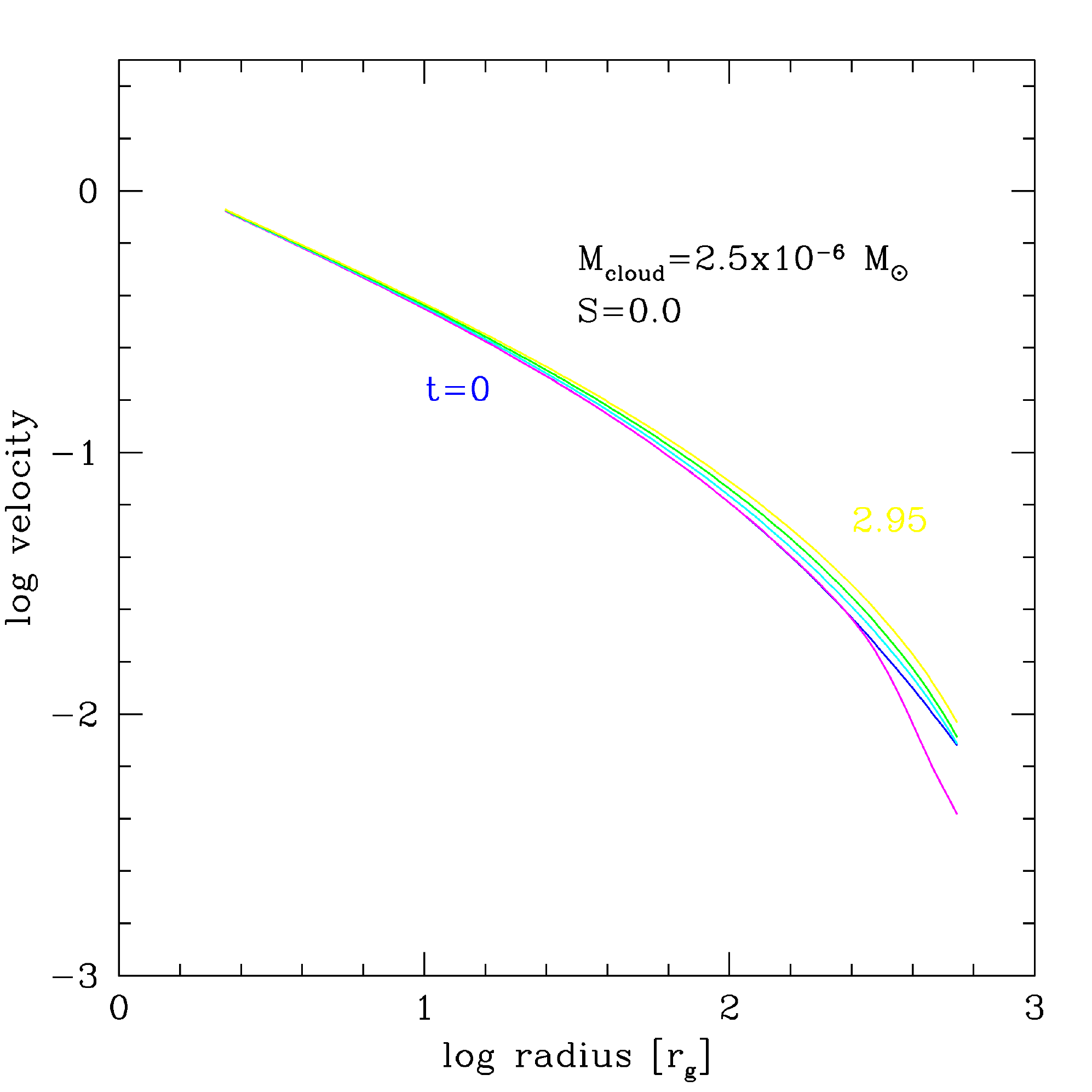}
  \includegraphics[width=7cm]{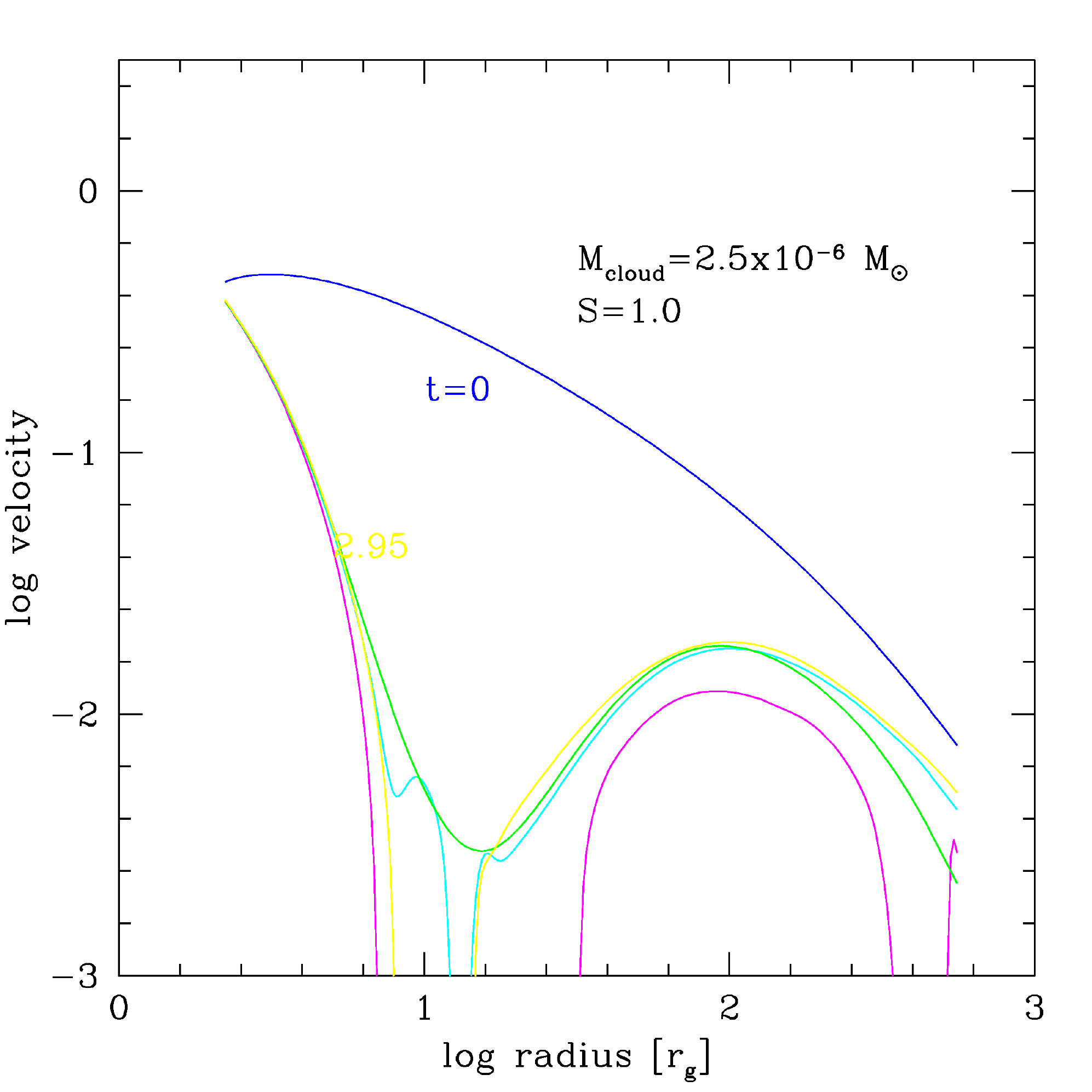}
  \caption{Density and velocity  as a function of time, for a very light
    cloud accreting onto a $3 M_{\odot}$ mass black hole. The label $S=0.0$ denotes solutions for a non-rotating flow, while $S=1.0$ involves slow rotation, imposed with the specific angular momentum as defined in Eq. (11). Several lines ion each plot refer to several snapshots in time, starting from $t=0$, until $t=2.95$ s.}
  \label{fig:profiles_light}
\end{figure}

In Figure \ref{fig:profiles_light}
we show the radial profiles of density and velocity in the cloud, for several time-snapshots during the simulation.
In the left panels, it can be seen that the non-rotating flow conserves our initial transonic Bondi
solution for the radial velocity.
The density profile also has the same slope for all snapshots, however its normalisation decreases, as the cloud very slowly empties (cf. Fig. \ref{fig:mdot_light}).
The sonic point was initailly located at $80 r_{\rm g}$, and it moved outwards only slightly (at the end of simulation, $t=2.95$ s, it shifted to  $\sim 120 r_{\rm g}$). This is because of the systematic decrease of density was not completely compensated by the decrease of pressure, since there is no matter supply from the outer boundary.

For comparison, we show here also the rotating 'Light Cloud' simulation. This model produced a local excess in
the density profile, above the Bondi solution, because the matter is slowed down by rotation and accumulates. This part of the flow extends only up to about $100 r_{\rm g}$. The flow here becomes almost completely subsonic in the equatorial plane, and the Mach number is equal to 1 only very close to the black hole ($r_{\rm s}=\sim 3 r_{\rm g}$).

\section{Pure Spherical test}

\begin{figure}
  \includegraphics[width=7cm]{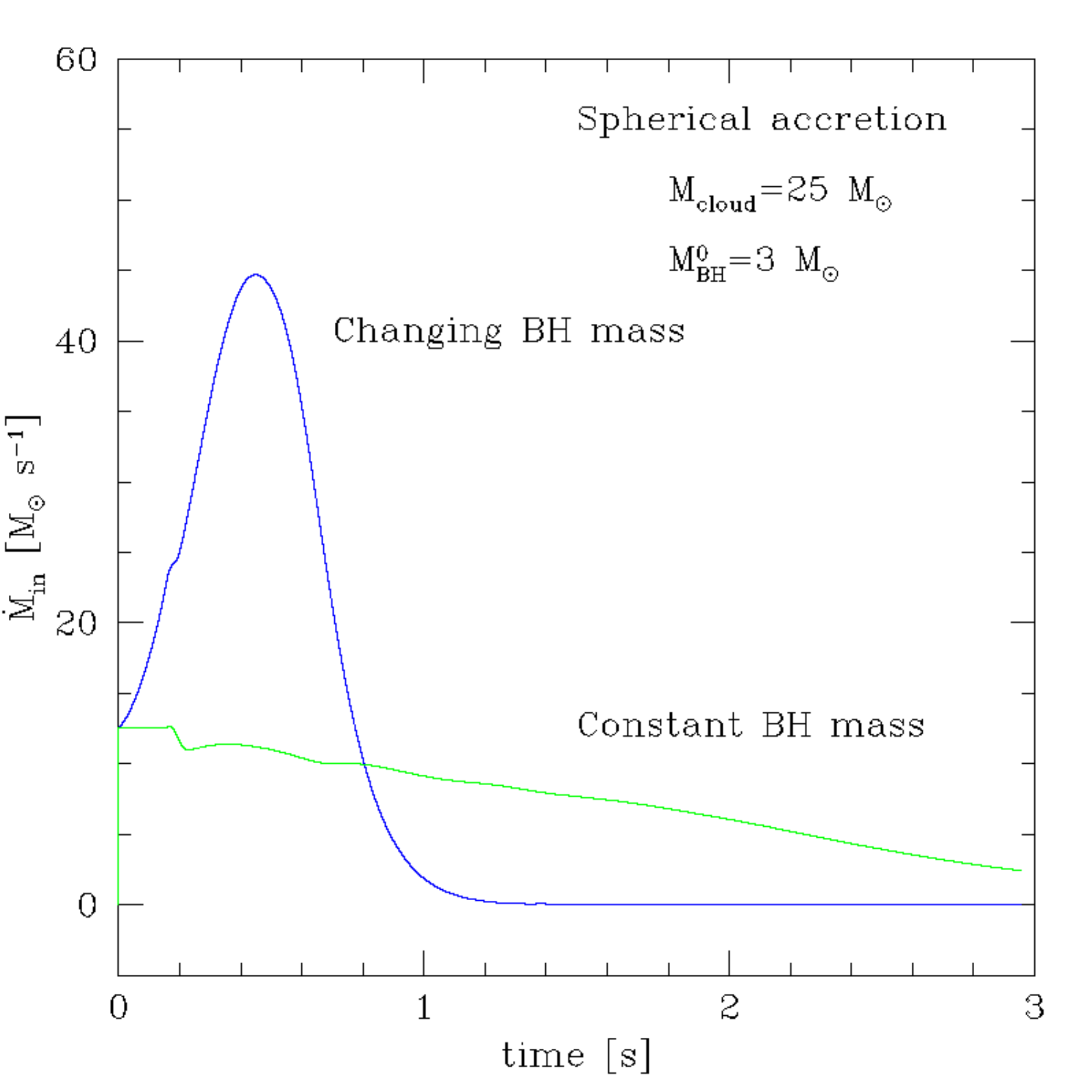}
  \includegraphics[width=7cm]{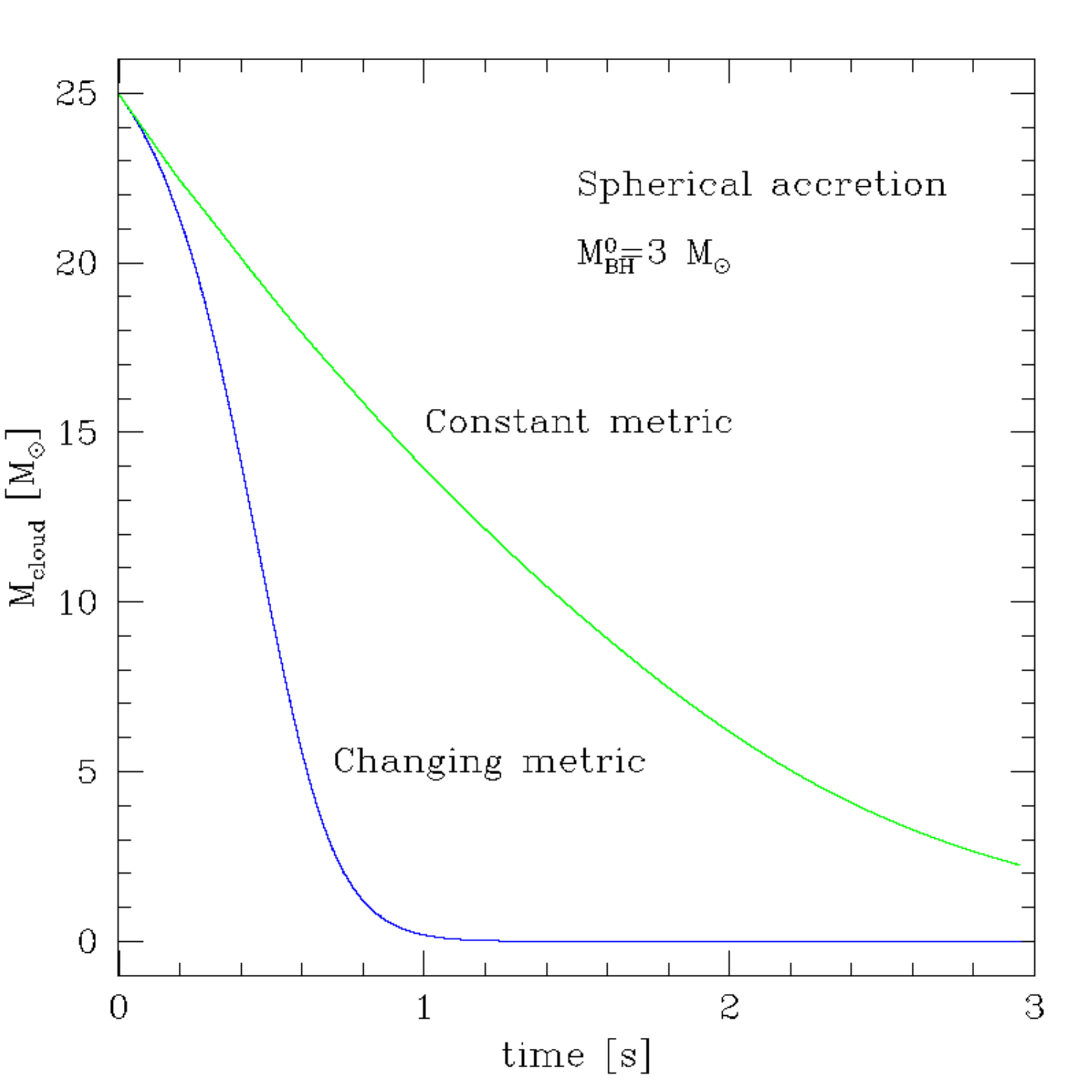}
  \caption{Mass accretion rate as a function of time, for a heavy
    cloud accreting onto a $3 M_{\odot}$ mass black hole. The model assumes no rotation. The static and changing metric solutions are different, as shown with green and blue lines, respectively. Right panel shows the corresponding total, volume integrated, mass of the cloud as it decreases in time.}
  \label{fig:mdot_puresph}
\end{figure}

In this test, we analyze the purely spherical simulation, when
the mass of the black hole grows very fast, as imposed by the large ratio of the accreting cloud mass to the mass of BH. We verify that the solution in general keeps the properties of the Bondi flow, regarding the profiles of density and radial velocity, as before.
We checked, that also in this test the spin of the black hole is constant and equal to zero. However, the BH mass grows significantly, while a fast decrease of the cloud's mass occurs. This is illustrated in Fig. \ref{fig:mdot_puresph}.

The figure shows for comparison the results for the code with the metric update routine disactivated. In this case, the mass accretion rate stays is small at the beginning of the simulation, and gradually decreases. It cannot be constant, because of the lack of matter supply from the outer boundary.
In contrast, the update of BH mass and metric change results in a huge mass accretion rate at the beginning of the simulation. This is because the
gravitational attraction of growing black hole pulls the matter much faster.

\begin{figure}
   \includegraphics[width=7cm]{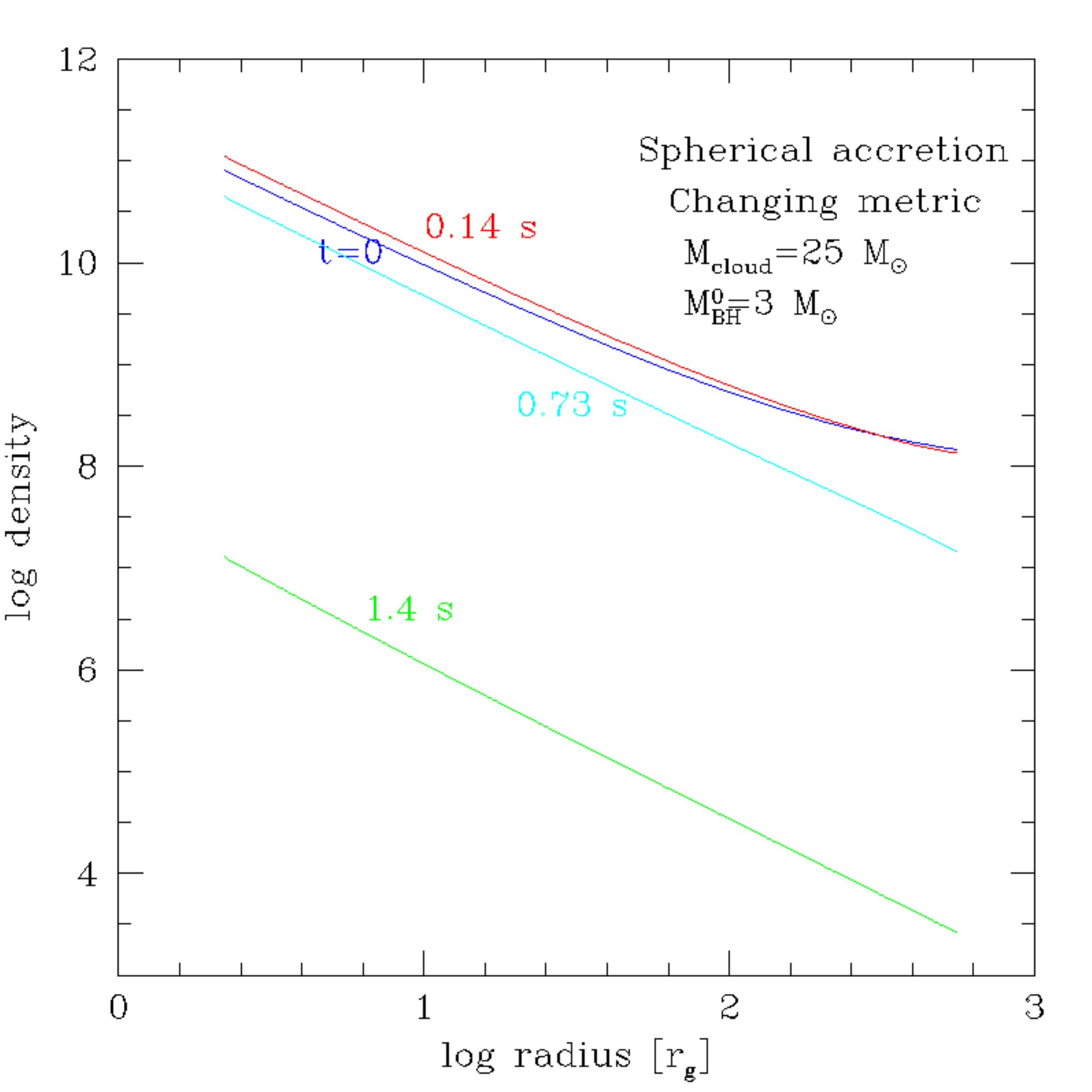}
   \includegraphics[width=7cm]{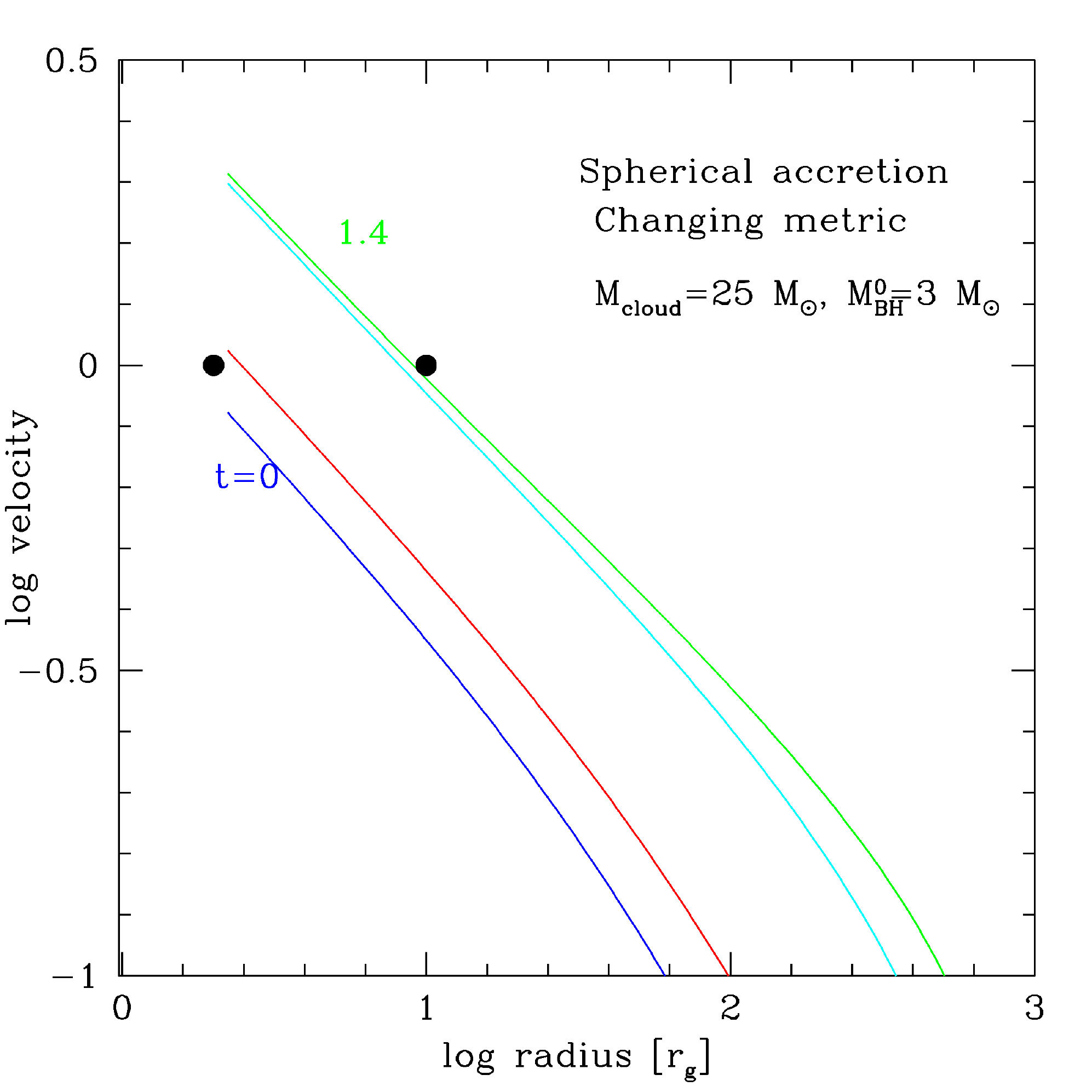}
  \caption{Density and velocity  as a function of time, for a heavy
    cloud accreting onto a $3 M_{\odot}$ mass black hole. The test neglects
    rotation, and BH spin is zero, while its mass grows and the space-time metric is updated accordingly. Several lines on each plot refer to several snapshots in time, starting from $t=0$, until $t=1.4$ s. The two black points on the right panel denote the location of the BH horizon, as implied by its starting and final mass value}
  \label{fig:profiles_puresph}
\end{figure}

In Figure \ref{fig:profiles_puresph} we show the profiles of density and
velocity for several time snapshots, in the simulation with changing metric.
(We note that the results for a constant metric are conserving the solutions from the initial condition and are very similar to the non-rotating Light Cloud test).
As shown in the left panel of Fig. \ref{fig:profiles_puresph}, the slopes of density profiles are constant for the first $\sim 0.5$ seconds, and the flow keep staying transonic, however the BH mass growth is very dynamical. Later on, the cloud is emptying and accretes very slowly onto the black hole, whose mass is already constant. The Bondi profile of density is reproduced, but for the purely supersonic flow.
The position of the sonic point is not constant, because it depends on both mass of the black hole, and on the sound speed at infinity.
After the first $\sim 0.15$ seconds, the BH mass grows by about 1.5 times, while the sonic point shifted from $80 r_{\rm g}$ to $140 r_{\rm g}$, i.e. by 1.75 times.
However, at the end of the simulation, when the mass of black hole increased 
by a factor of 5, the sonic point reached almost the outer boundary of the grid, and was located at radius more than 11 times larger than initially. This fact can be explained as the response of the accreting matter to the growing mass of the 
black hole and the metric change. Both density and pressure have to adjust to the new conditions.

The profiles of velocity conserve their constant slopes, same as in the initial condition, while the magnitude of velocity systematically increase.
It can be noted, that the apparently super-luminal velocities are
reached below the current black hole horizon ($r_{\rm H}=2 M$ for a Schwarzschild black hole).
It shows therefore, that the black hole mass growth by about five times (from $3 M_{\odot}$ up to $\sim 15 M_{\odot}$) resulted in the proportional
growth of the horizon size, as marked by the black points in the right panel of the Figure \ref{fig:profiles_puresph}. We conclude that this test confirms
the consistency of our simulation, because the radius of the BH horizon, in contrast to the sonic radius, should depend only on the black hole mass.


\begin{thebibliography}{0} 

\bibitem[Abbott et al. (2016)]{abbott2016} Abbott, B.-P., et al., 2016, {\it Phys. Rev. Lett.}, {\bf 116}, 6

\bibitem[Barkov \& Komissarov (2008)]{bkomis08} Barkov, M.V., Komissarov S.S., 
2008, {\it MNRAS}, {\bf 385}, L28

\bibitem[Barkov \& Komissarov (2010)]{bkomis10} Barkov, M.V., Komissarov S.S., 
  2010, {\it MNRAS}, {\bf 401}, 1644

\bibitem[Batta et al. (2017)]{batta2017} Batta, A., Ramirez-Ruiz, E., Fryer, C., 2017, {\it MNRAS}, {\bf 846}, 15
  
\bibitem[Batta \& Lee (2016)]{batta2016} Batta, A., Lee, W.H., 
2016, {\it MNRAS}, {\bf 459}, 2140 

\bibitem[Crowther (2007)]{crowther2007} Crowther, P.A., 2007, {\it A\&A Rev.}, {\bf }, 45, 177
  
\bibitem[Chakrabarti \& Das (2001)]{chakrabarti2001} Chakrabarti, S., Das, S., 2001, {\it MNRAS}, {\bf 327}, 808

\bibitem[Das (2002)]{das2002} Das, T., 2002, {\it ApJ}, 577, 880
  
\bibitem[Culter \& Flanagan (1994)]{cutler1994} Cutler, C., Flanagan, E.E., 1994, {\it Phys. Rev. D}, {\bf 49}, 2658

\bibitem[Gammie et al. (2004)]{gammie2004} Gammie C.F., McKinney J.C. \& Shapiro, 2004, {\it ApJ}, {\bf 602}, 312 

\bibitem[Gammie et al. (2003)]{gammie2003harm} Gammie C.F., McKinney J.C. \& Toth G., 2003, {\it ApJ}, {\bf 589}, 444 

\bibitem[Hamersky \& Karas (2008)]{karas} Hamersky, J., Karas, V., 2013, {\it A\&A}, 555, 32

\bibitem[Janiuk \& Proga (2008)]{janiukproga2008} Janiuk A., Proga D., 2008, {\it ApJ}, {\bf 675}, 519

\bibitem[Janiuk, Moderski \& Proga (2008)]{moderski} Janiuk A., Moderski R., Proga D., 2008, {\it ApJ}, {\bf 687}, 433 

\bibitem[Janiuk et al. (2013)]{bbh2013} Janiuk A., Charzynski S., Bejger M., 2013, {Astronomy \& Astrophysics}, {\bf 560}, A25 

\bibitem[Janiuk et al. (2013)]{janiuketal2013} Janiuk A., Mioduszewski P., Moscibrodzka M., 2013, {\it ApJ}, {\bf 776}, 105

\bibitem[Janiuk (2017)]{janiuk2017} Janiuk A., 2017, {\it ApJ}, {\bf 837}, 39 

\bibitem[Janiuk et al. (2017)]{newa2017} Janiuk A., Bejger M., Charzynski S., Sukova P., 2017, {\it New Astronomy}, {\bf 51}, 7

\bibitem[Kastaun et al. (2017)]{kastaun} Kastaun W., Ciolfi R., Endrizzi A., Giacomazzo B., 2017, {\it Phys Rev D.}, {\bf 96}, 043019

\bibitem[Kuroda et al. (2018)]{kuroda2018} Kuroda T., Kotake K., Takiwaki T., Thielemann F.-K., 2018, {\it MNRAS}, {\bf 477}, L80
  
\bibitem[Mach, Pirog \& Font (2018)]{mach2018} Mach P., Pirog M., Font, J., 2018, {\it Class. Q. G.}, {\bf 35}, 095005

\bibitem[McKinney \& Gammie (2004)]{McKinneyGammie2004} McKinney J.C., Gammie C.F., 2004, {\it ApJ}, {\bf 611}, 977

\bibitem[Noble et al. (2006)]{noble2006} Noble S.C., Gammie C.F., McKinney J.C., \& Del Zanna L., 2006, {\it ApJ}, {\bf 641}, 626

\bibitem[Lentz et al. (2015)]{lentz2015} Lentz E.J., 2015, {\it ApJL}, {\bf 807}, 31
  
\bibitem[Liu et al. (2015)]{liu2015} Liu, T., Hou, S.-J., Xue, L., Gu, W.-M., 2015, {\it ApJS}, {\bf 218}, 12
  
\bibitem[Lopez-Camara et al. (2010)]{lopez2010} Lopez-Camara, D., Lee W.H., Ramirez-Ruiz E., 2010, {\it ApJ}, {\bf 716}, 1308

\bibitem[Ott et al. (2018)]{ott2018} Ott C.D., et al., 2018, {\it ApJL}, {\bf 855}, 3
  
\bibitem[Paczynski (1998)]{paczynski98} Paczynski B., 1998, {\it ApJL}, {\bf 494}, 45
  
\bibitem[Piran (2004)]{piran2004} Piran T., 2004, {\it Rev. Mod. Phys}, {\bf 76}, 1143 

\bibitem[Pankov et al. (2017)]{pankov17} Pankov C., et al., 2017, {\it ApJ}, {\bf 834}, 154
  
\bibitem[Podsiadlowski et al. (2004)]{podsiadlowski04} Podsiadlowski, P., et al., 2004, {\it ApJ}, {\bf 607}, 17

\bibitem[Spera et al. (2015)]{spera2015} Spera, M., Mapelli, M., Bressan, A., 2015, {\it MNRAS}, {\bf 451}, 4086

\bibitem[Semerak \& Sukova (2010)]{semerak} Semerak O., Sukova P., 2010, {\it MNRAS}, {\bf 404}, 545
  
\bibitem[Sukova \& Janiuk (2015)]{sukova2015} Sukova P., Janiuk A., 2015, {\it MNRAS}, {\bf 447}, 1565
  
\bibitem[Sukova et al. (2017)]{sukova2017} Sukova P., Charzynski S., Janiuk A., 2017, {\it MNRAS}, {\bf 472}, 4327 

\bibitem[Takahashi et al. (2014)]{takahashi2014} Takahashi K., Umeda H., Yoshida T., 2014, {\it ApJ}, {\bf 794}, 40
  
\bibitem[Woosley (1993)]{woosley1993} Woosley S.E., 1993, {\it ApJ}, {\bf 405}, 273 

\bibitem[Woosley et al. (2002)]{woosley2002} Woosley S.E., Heger A., Weaver T.A., 2002, {\it Rev. Mod. Phys.}, {\bf 74}, 1015 

\end{thebibliography}
\end{document}